\documentclass[sigconf]{acmart}
\AtBeginDocument{%
  }

\setcopyright{cc}
\setcctype{by-nc-nd}
\copyrightyear{2026}
\acmYear{2026}
\acmDOI{}
\acmConference[MM '26]{the 34th ACM International Conference on Multimedia}
  {November 10--14, 2026}{Rio de Janeiro, Brazil}
\acmISBN{979-8-4007-2213-4/2026/11}

\usepackage{graphicx}

\usepackage{booktabs} \usepackage{multirow} \usepackage{graphicx}

\usepackage[most]{tcolorbox}
\usepackage{xcolor}
\usepackage{enumitem}
\usepackage{tabularx}
\usepackage{fvextra}

\setlist[itemize]{leftmargin=1.15em, topsep=2pt, itemsep=1pt, parsep=0pt, partopsep=0pt}

\tcbset{
  promptbox/.style={
    enhanced,
    width=\columnwidth,
    colback=white,          
    colframe=black!45,      
    colbacktitle=black!12,  
    coltitle=black,         
    fonttitle=\bfseries,    
    boxrule=0.8pt,
    titlerule=0.5pt,        
    arc=0pt,                
    left=6pt, right=6pt,
    top=8pt, bottom=6pt,    
    toptitle=3pt,           
    bottomtitle=3pt,        
    title={#1}              
  },
  codebox/.style={
    enhanced,
    colback=black!3,
    colframe=black!18,
    boxrule=0.5pt,
    arc=0pt,
    left=5pt, right=5pt,
    top=4pt, bottom=4pt,
  }
}

\begin{document}

\title{Unsupervised Multimodal Intent Discovery via MLLM-Guided Concept Generation and Semantic Propagation}


\author{Yunjin Gu}
\affiliation{%
  \institution{The Chinese University of Hong Kong, Shenzhen}
  \city{Shenzhen}
  \country{China}}
\email{yunjingu@link.cuhk.edu.cn}

\author{Qianrui Zhou}
\affiliation{%
  \institution{Tsinghua University}
  \city{Beijing}
  \country{China}
}
\email{zgr22@mails.tsinghua.edu.cn}

\author{Hua Xu}
\affiliation{%
 \institution{Tsinghua University}
 \city{Beijing}
 \country{China}}
\email{xuhua@tsinghua.edu.cn}

\renewcommand{\shortauthors}{Trovato et al.}

\begin{abstract}
  Unsupervised multimodal intent discovery aims to uncover latent intents from unlabeled multimodal dialogues, but remains challenging due to the lack of explicit semantic supervision. Existing methods often provide limited interpretability, as their refinement mainly relies on geometric similarity rather than high-level semantic guidance.
To address these limitations, we propose \textbf{MCSP}, a fully unsupervised method that introduces semantic refinement based on concepts into multimodal intent discovery.
To obtain reliable semantic evidence for intent discovery, we identify high-quality representative samples for each cluster and use them to support MLLM-guided contrastive reasoning against neighboring clusters, which produces interpretable high-level semantic concepts. Building on these concepts, we perform semantic propagation over a semantically weighted graph to align conceptual information with local structural consistency and generate reliable pseudo-labels for representation refinement.
Extensive experiments on three challenging multimodal intent datasets show that MCSP consistently outperforms state-of-the-art methods while producing interpretable clusters grounded in semantic concepts.
\end{abstract}


\maketitle

\section{Introduction}
Unsupervised multimodal intent discovery has emerged as a pivotal task in multimodal dialogues, facilitating the identification of latent semantic structures while eliminating the need for costly manual labeling.
This unlabeled approach demonstrates significant utility in various domains, including video content recommendation, efficient multimodal data annotation, and virtual human technologies~\citep{zhang-etal-2024-unsupervised}. 

With the rise of multimodal analysis, understanding intent within conversations has been extensively explored. This field is anchored by diverse benchmarks, ranging from EMOTyDA \citep{Saha-etal-2020-towards}, which provides multi-party dialogue segments with aligned textual, acoustic, and visual modalities, to the MIntRec series \citep{zhang-etal-2022-mintrec, zhang-etal-2024-mintrec2}, which offers specialized annotations across 20 and 30 categories of high granularity. The availability of these resources has driven the evolution of models that achieve superior performance, such as TCL-MAP \citep{zhou-etal-2024-token-level}, DuoDN \citep{chen-etal-2024-dual-oriented}, and InMu-Net \citep{zhu-etal-2024-inmu-net}.


\begin{figure}[t]
  \centering
  \includegraphics[width=\linewidth]{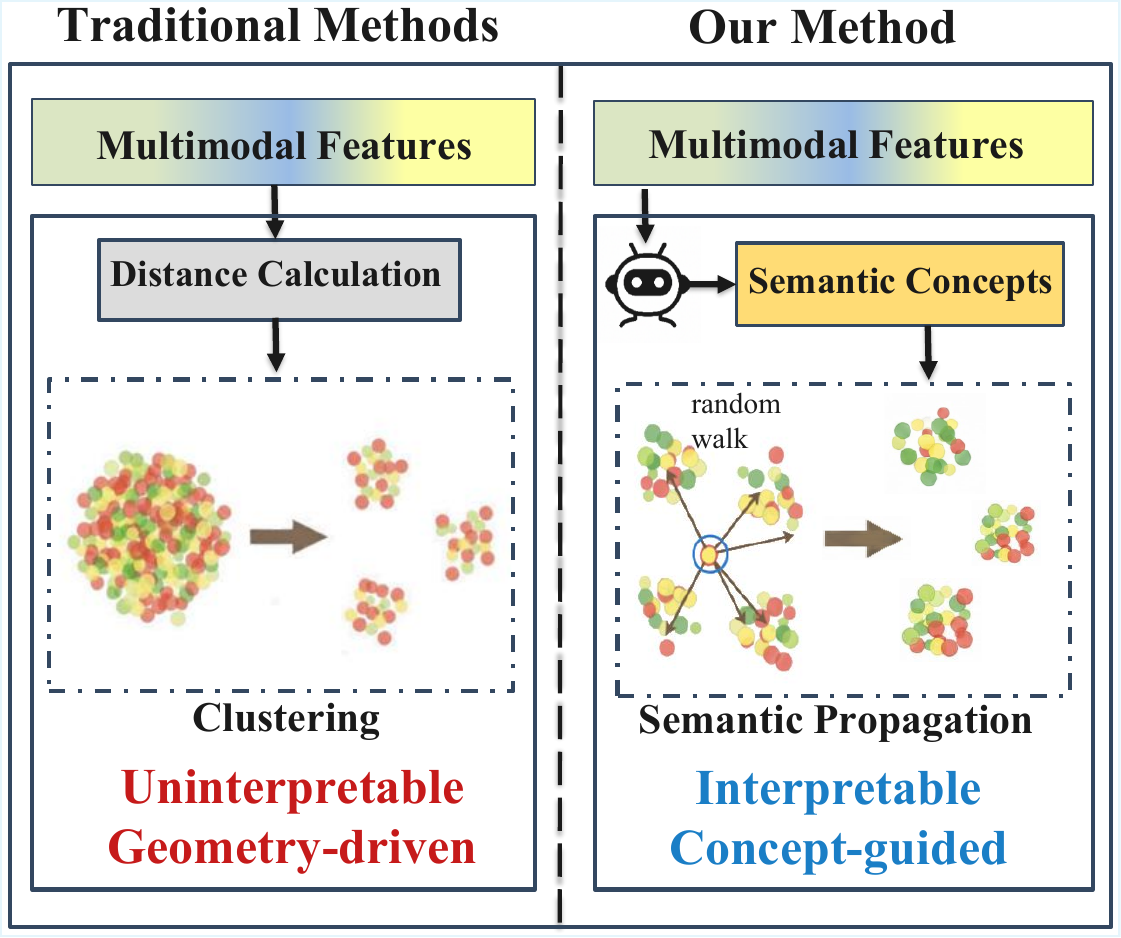}
  \caption{Comparison of intent discovery paradigms.}
  \Description{A schematic comparison of intent discovery paradigms, illustrating the limitations of prior methods and the key idea of the proposed concept-guided method.}
  \label{fig:intro_concept_guided}
\end{figure}

However, these methods depend heavily on predefined labels, making them impractical for discovery tasks where intent categories are unknown. 
Consequently, intent discovery has emerged as a vital paradigm, typically treated as a clustering task under partial or no label supervision. Early efforts focused on the classic clustering of embeddings followed by post hoc summarization \citep{cheung-li-2012-sequence,padmasundari-bangalore-2018-intent,haponchyk-etal-2018-supervised}. Recent progress has shifted toward integrating representation learning with iterative clustering, where pseudo labels are utilized to refine the structure of clusters \citep{zhang-etal-2022-new}. Furthermore, strategies that employ prompts leverage the semantics of labels as priors to maintain coherence \citep{liang-liao-2023-clusterprompt}. Recent LLM-guided methods advance intent discovery through instance selection, textual centroids, and keyphrase-guided prototypes, yet fall short of multimodal concept discovery \citep{lin-etal-2025-spill,diaz-rodriguez2026summaries,kim-etal-2025-kstc}. Moving beyond unimodal paradigms, UMC \citep{zhang-etal-2024-unsupervised} leverages contrastive learning to forge integrated multimodal representations. However, as illustrated in the left panel of Figure \ref{fig:intro_concept_guided}, traditional clustering methods suffer from two limitations. 
First, existing approaches still lack high-level, interpretable concepts from multimodal data to guide clustering.
Second, they rely heavily on the geometric properties of the data, 
which fails to capture the underlying semantic consistency required for accurate discovery.

To address these challenges, we propose MCSP, a method for Multimodal Concept-Guided Semantic Propagation that shifts the discovery paradigm from a purely geometric search into a process of discovery anchored by interpretable semantic concepts, as shown in the right panel of Figure \ref{fig:intro_concept_guided}. Our method first introduces a progressive refinement strategy to identify high-quality representative samples, effectively insulating the model from the instability of ambiguous boundary instances. These representatives then serve as the empirical foundation for an MLLM-guided reasoning process. By leveraging the inferential capacity of an MLLM, we perform a contrastive reasoning across target and neighboring clusters to generate explicit concepts. These concepts act as global semantic anchors, bridging the gap between abstract feature distributions and explicit semantic information. 

To integrate high-level priors with the underlying data distribution, we introduce a semantic propagation mechanism grounded in a concept-aware neighborhood graph. Instead of relying on simple proximity, our approach calibrates edge weights through multimodal semantic alignment, ensuring that feature diffusion is constrained within coherent conceptual boundaries. By initiating iterative random walks \cite{zhu-etal-2003-Learning} from high-quality representatives, the system allows semantic information to flow across the manifold, harmonizing global conceptual guidance with local structural consistency. Finally, the system is optimized via a concept-supervised contrastive learning loss, utilizing multi-view augmentations to enforce intra-concept compactness and inter-concept separation. This establishes a synergistic feedback loop where the embedding manifold and the conceptual graph mutually reinforce one another, achieving a reliable and transparent identification of intents.

We summarize our contributions as follows. (1) For fully unsupervised discovery, we introduce a method that leverages MLLM reasoning over high-quality representatives to generate high-level semantic concepts. (2) We propose a semantic propagation mechanism that diffuses conceptual evidence across a concept-aware graph by initiating random walk from trusted landmarks, effectively harmonizing global semantic priors with the local manifold structure of the data. (3) Extensive experiments validate that MCSP consistently outperforms existing methods while generating interpretable semantic concepts
that justify the discovery of intents.

\section{Related Work}


\label{related}

\begin{figure*}[t]
  \centering
  \includegraphics[width=\textwidth]{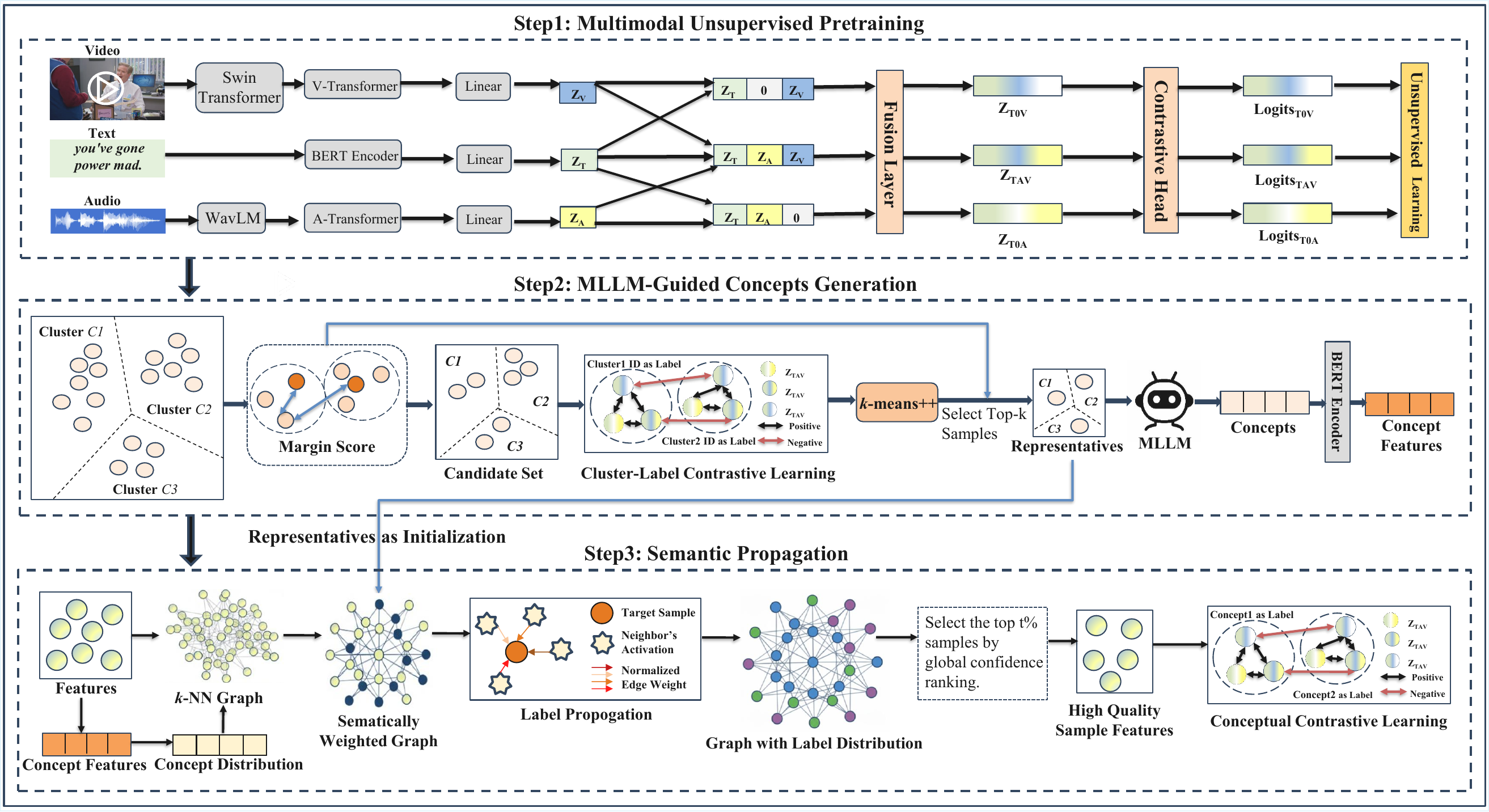}
  \caption{Overview of the proposed MCSP method. The MCSP proceeds through three synergistic stages to achieve precise unsupervised intent discovery: (1) Contrastive Pretraining, which performs mask-based contrastive learning on multimodal inputs to establish a robust joint embedding space; (2) Concept Extraction, which leverages MLLM-guided reasoning to generate semantic concepts from high-quality representative samples; and (3) Semantic Propagation, which integrates evidence diffusion and concept-driven supervision to refine the embedding manifold for accurate intent identification.}
  \Description{A pipeline overview of the proposed MCSP framework for unsupervised intent discovery. The method contains three stages: contrastive pretraining on multimodal inputs, semantic concept extraction from representative samples using MLLM-guided reasoning, and semantic propagation that combines evidence diffusion with concept-driven supervision to refine the embedding space for intent identification.}
  \label{fig:method_overview}
\end{figure*}

\subsection{Unsupervised Clustering}

Unsupervised clustering has transitioned from classical similarity based paradigms \citep{macqueen-1967-some, gowda-krishna-1978-agglomerative} toward deep learning frameworks that prioritize the joint optimization of representation learning and cluster assignment. Specifically, early models such as DEC \citep{xie-etal-2016-unsupervised} and DCN \citep{yang-etal-2017-towards} utilized iterative refinement, while DeepCluster \citep{caron-etal-2018-deep} employed clustering as a pretext task for large scale feature learning. Recently, contrastive clustering frameworks \citep{zhang-etal-2021-supporting, li-etal-2021-contrastive} have integrated instance level invariance with cluster level refinement to produce clustering friendly representations. While LLM-guided clustering has emerged to provide semantic refinement, these methods are designed for text and face high query complexity as detailed in Appendix \ref{app:llm_clustering}. 
Specifically, SPILL \citep{lin-etal-2025-spill} employs selection and pooling mechanisms to leverage LLM-driven feature refinement for domain-adaptive clustering while remaining restricted to the textual modality.
Expanding beyond single modality data, recent research focuses on capturing semantic correlations across multimodal signals. While earlier efforts \citep{chen-etal-2021-multimodal} introduced cross modal constraints, they struggled with fine grained conversational semantics. To address this, \citet{zhang-etal-2024-unsupervised} established UMC, the first unsupervised framework for multimodal intent discovery, which leverages multimodal augmentation and density based sample selection to facilitate clustering.

\subsection{Intent Discovery}
Intent discovery seeks to extract latent categories from unlabeled data. Early efforts relied on weak auxiliary or lightweight supervision to facilitate clustering, but failed to capture robust, high-level semantics in large-scale applications \citep{hakkani-tur-etal-2015-clustering, haponchyk-etal-2018-supervised}. This led to a paradigm shift toward semi-supervised discovery, where labeled intents serve as anchors to align and cluster novel categories \citep{zhang-etal-2022-new, mou-etal-2022-watch, shi-etal-2023-diffusion}. However, the heavy reliance on labels limits their utility in purely unsupervised scenarios. To bridge this gap, USNID \citep{zhang-etal-2024-clustering} introduced an unsupervised centroid-guided refinement pipeline, yet remains confined to the textual modality. UMC \citep{zhang-etal-2024-unsupervised} advances multimodal discovery as a leading baseline, effectively unifying tri modal signals via modality masked contrastive learning. Appendix \ref{app:openworld_intent} provides further context on open world intent discovery.

\section{Methodologies}

In this section, we introduce the detail of our
proposed method, MCSP, as illustrated in Figure \ref{fig:method_overview}.

\subsection{Multimodal Unsupervised Pre-training}
\label{sec:pretraining}
MCSP adopts the architecture and pre-training of UMC \citep{zhang-etal-2024-unsupervised} to initialize a representation space that captures latent cluster structures.



For each instance $s_i = (s_i^{T}, s_i^{A}, s_i^{V})$, we obtain a unified representation by projecting modality-specific features into a shared latent space. Textual representations $z_i^{T} \in \mathbb{R}^{D_H}$ are obtained by fine-tuning BERT \cite{devlin-etal-2019-bert} and projecting the $[\mathrm{CLS}]$ embedding through a linear layer $f_T(\cdot)$. For non-verbal modalities, we use Swin Transformer \cite{liu-etal-2021-swin} and WavLM \cite{chen-etal-2022-wavlm} as high-capacity encoders to extract sequence features. These sequences are further processed by modality-specific Transformers \cite{vaswani-etal-2017-attention} to generate global summaries $z_i^{A}$ and $z_i^{V}$ \cite{tsai-etal-2019-multimodal}. Finally, these embeddings are integrated through a non-linear fusion layer $F(\cdot)$ to model cross-modal interactions:
\begin{equation}
\mathbf{z}_i^{TAV} = F\left(\mathrm{Concat}(\mathbf{z}_i^{T}, \mathbf{z}_i^{A}, \mathbf{z}_i^{V})\right) \in \mathbb{R}^{D_H}.
\label{eq:pretrain_fuse}
\end{equation}
Following \cite{zhang-etal-2024-unsupervised}, we strengthen robustness by enforcing cross-modal consistency across views generated via selective audio-visual masking. Optimized with an InfoNCE objective, this process regularizes the latent space through instance-level contrastive alignment, providing a stable foundation for subsequent intent discovery.
\subsection{MLLM-Guided Concepts Generation}
\label{sec:concept_anchors}

While unsupervised pre-training organizes embedding spaces geometrically, such representations often suffer from semantic opacity and cluster drift leading to noisy or misaligned groupings.
To address this limitation, MCSP introduces MLLM-Guided concepts as global semantic anchors to ground latent clusters in explicit conceptual taxonomies. 
This process unfolds in two phases: High-Quality Representatives Selection and Semantic Concept Generation.


To ensure stable and grounded concept generation, we prioritize selecting a compact set of high-quality representative samples that reside in the unambiguous regions of the embedding space. Theoretically, samples located near cluster boundaries are prone to semantic noise and inter-cluster interference in unsupervised scenarios. To mitigate this, we introduce a margin-based filtering mechanism to identify instances with high cluster-membership certainty. Specifically, we first perform $k$-means++ \cite{Arthur-etal-2007-Kmeans++} on the fused embeddings $\{\mathbf{z}_i\}_{i=1}^{N}$ to obtain cluster assignments $a_i \in \{1,\dots,K\}$ and centroids $\{\boldsymbol{\mu}_k\}_{k=1}^{K}$. For each instance $i$, a margin score $m_i$ is computed to quantify its representativeness and discriminative power:
\begin{equation}
m_i = \min_{k \neq a_i} d(\mathbf{z}_i, \boldsymbol{\mu}_k) - d(\mathbf{z}_i, \boldsymbol{\mu}_{a_i}),
\label{eq:margin_global}
\end{equation}
where $d(\cdot,\cdot)$ denotes the Euclidean distance. Intuitively, a larger $m_i$ indicates that the sample is deeply embedded within its assigned cluster and well-separated from potential competitors. As further validated by the comparative analysis in Appendix \ref{sec_score}, this margin-based criterion effectively isolates the top $m$ prototypical instances within each cluster to form the high-confidence candidate set $\mathcal{C}$.

To enhance the semantic coherence, we perform supervised contrastive learning over $\mathcal{C}$ using the assignments $\{a_i\}_{i \in \mathcal{C}}$ obtained from the initial clustering as pseudo-labels. This optimization objective enforces intra-cluster compactness and inter-cluster separation, yielding a more discriminative manifold $\{\mathbf{z}'_i\}_{i=1}^{N}$.

The final representative set $\mathcal{S}$ is identified by applying $k$-means++ for a second time to the refined embeddings $\{\mathbf{z}'_i \mid i \in \mathcal{C}\}$. By evaluating the samples in this discriminative manifold, we obtain updated centroids $\{\boldsymbol{\mu}'_k\}_{k=1}^{K}$ and refined margin scores $m'_i$. The final representatives for each cluster $k$ are then selected as:
\begin{equation}
\mathcal{S}_k = \{ i \in \mathcal{C} \mid a'_i = k \}_{1:N_r},
\label{eq:seed_set}
\end{equation}
where $\{ \cdot \}_{1:N_r}$ denotes the subset of $N_r$ instances ranked in descending order of their refined margin scores $m'_i$, and $a'_i$ represents the updated cluster assignment. 
By prioritizing prototypical instances, this progressive refinement ensures robust concept generation and effectively filters out noisy samples from overlapping regions.

\begin{figure}[t]
  \centering
  \includegraphics[width=\linewidth]{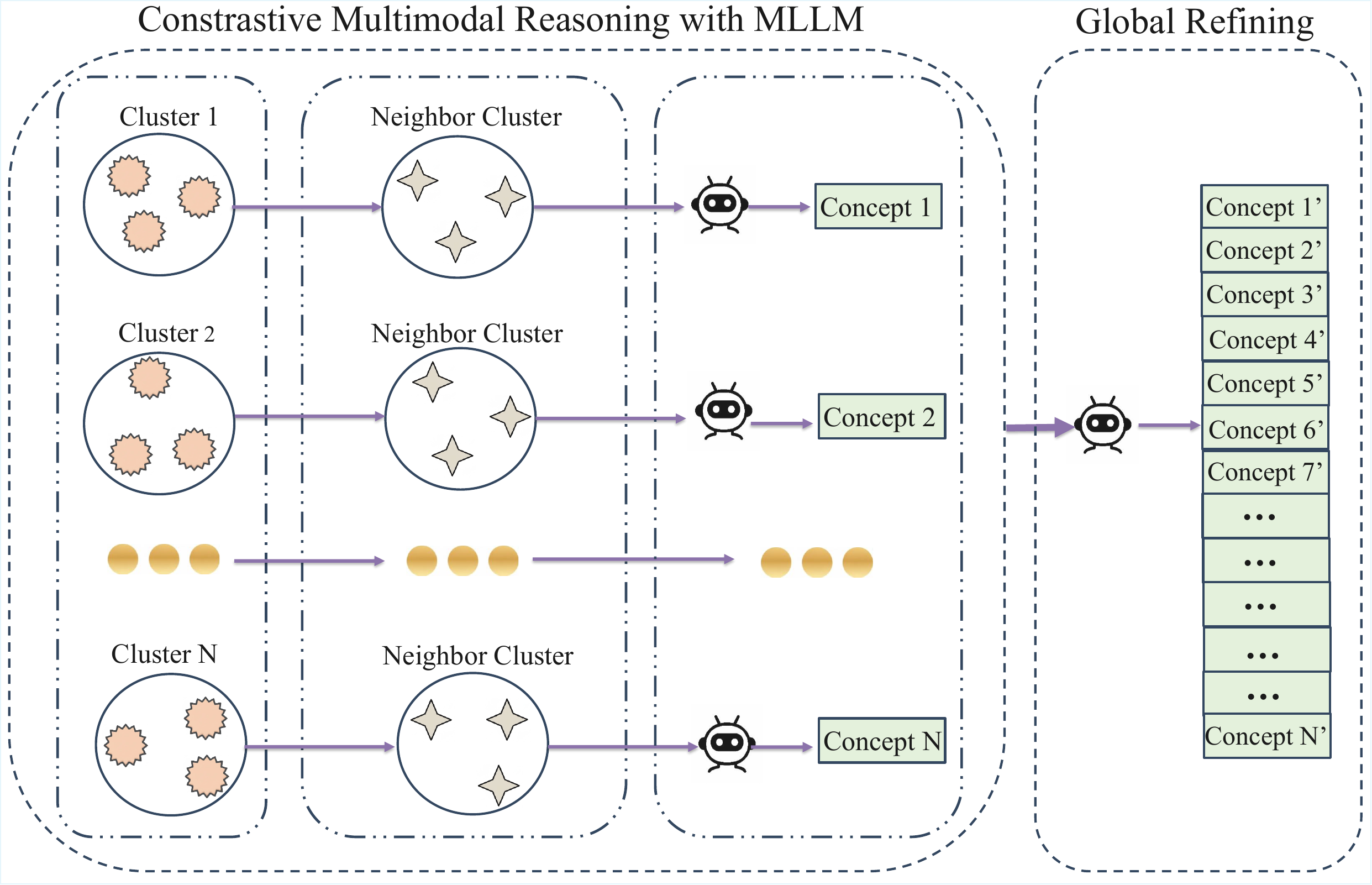}
  \caption{Multimodal Contrastive Reasoning.}
  \Description{An illustration of multimodal contrastive reasoning for concept generation, showing how the model uses multimodal inputs and comparative reasoning to derive semantic concepts.}
  \label{fig:concept_anchoring}
\end{figure}

Building upon the curated representatives, we leverage advanced MLLMs, including Gemini-3.0-pro and Qwen3-VL, as discriminative reasoners to extract semantic anchors via a multimodal contrastive strategy,  as shown in Fig.~\ref{fig:concept_anchoring}. Specifically, for each cluster $k$, the model is presented with the multimodal evidence from target representatives $i \in \mathcal{S}_k$ alongside that of its nearest neighboring cluster $j \in \mathcal{S}_{\text{nn}(k)}$. Each representative is characterized by its respective video $\mathcal{V}$, which comprises both visual and audio information, and its associated text $\mathcal{T}$. By leveraging this enriched contextual information $\{\mathcal{V}_i, \mathcal{T}_i\}_{i \in \mathcal{S}_k}$ and $\{\mathcal{V}_j, \mathcal{T}_j\}_{j \in \mathcal{S}_{\text{nn}(k)}}$, the MLLM effectively captures the semantic discrepancies between clusters, enabling the generation of precise concepts. Formally, this contrastive reasoning process is expressed as:
\begin{equation}
t_k = \text{MLLM} \left( \{ \mathcal{V}_i, \mathcal{T}_i \}_{i \in \mathcal{S}_k \cup \mathcal{S}_{\text{nn}(k)}}; \text{Prompt}_{\text{cont}} \right),
\label{eq:contrastive_reasoning}
\end{equation}
where $t_k$ denotes the discriminative concept. A subsequent global refining stage reviews the initial set $\mathcal{T}_{\text{init}} = \{t_k\}_{k=1}^{K}$ collectively to ensure syntactic consistency and global coherence across the entire concept space. This refinement process is formulated as:
\begin{equation}
\mathcal{T}_{\text{refined}} = \text{MLLM}(\mathcal{T}_{\text{init}}; \text{Prompt}_{\text{global}}),
\label{eq:global_refining}
\end{equation}
where $\mathcal{T}_{\text{refined}} = \{t'_k\}_{k=1}^{K}$ denotes the set of refined phrases. The operation unifies the linguistic syntax and maintains concept consistency across the set, ensuring that each $t'_k$ serves as a standardized textual representation with a coherent style. Then, each refined concept $t'_k$ is encoded into a anchor $\mathbf{c}_k \in \mathbb{R}^{D_H}$. 
To maintain cross-modal consistency during text-only processing, $t'_k$ is passed through the pre-trained BERT and fusion modules while concurrently utilizing zero-masked audio and video streams to satisfy the multimodal architectural requirements.
Detailed specifications of $\text{Prompt}_{\text{cont}}$ and $\text{Prompt}_{\text{global}}$ are provided in Appendix \ref{sec:prompt_design} for further reference.

\subsection{Semantic Propagation}
\label{sec:concept_rep_learning}

Although concept anchors provide global references, direct nearest neighbor assignment neglects the underlying data manifold and potential concept feature misalignment, validated by the ablation of $k$-NN graph in Section \ref{method:graph_ablation}. 

To mitigate this, we regularize anchor semantics through local geometric smoothness by diffusing concept evidence over a neighborhood graph. We first represent the local structure as a $k$-NN graph $\mathcal{G} = \{\mathcal{V}, \mathcal{E}\}$, where $\mathcal{V}$ denotes the set of instances and $\mathcal{N}(i)$ represents the neighbor set containing the top $k$ samples for instance $i$. For each edge $(i, j)$ with $j \in \mathcal{N}(i)$, we define a geometric weight $w^{g}_{ij}$ via a temperature scaled exponential kernel:
\begin{equation}
w^{g}_{ij}= \frac{\exp\left(\mathrm{sim}(\mathbf{z}_i, \mathbf{z}_j) / \tau_g\right)}{\sum_{j' \in \mathcal{N}(i)} \exp\left(\mathrm{sim}(\mathbf{z}_i, \mathbf{z}_{j'}) / \tau_g\right)},
\label{eq:geom_weight}
\end{equation}
where $\tau_g$ is a temperature parameter controlling connectivity sharpness. To incorporate semantic constraints, we derive the concept distribution for each sample with respect to the semantic anchors:
\begin{equation}
P_{ik} = \frac{\exp(\mathrm{sim}(\mathbf{z}_i,\mathbf{c}_k)/\tau_c)}{\sum_{k'=1}^K \exp(\mathrm{sim}(\mathbf{z}_i,\mathbf{c}_{k'})/\tau_c)},
\label{eq:concept_posterior}
\end{equation}
where $\tau_c$ is a temperature scaling factor. Based on these distributions, we quantify the semantic consistency between neighboring samples $i$ and $j$ as $s^c_{ij}=\langle P_i, P_j\rangle$, where $P_i \in \mathbb{R}^K$ represents the posterior vector for sample $i$ and $\langle \cdot, \cdot \rangle$ denotes the inner product operator. The geometric weights are subsequently refined through semantic alignment:
\begin{equation}
\tilde{w}_{ij}=w^{g}_{ij}(1+\lambda s^c_{ij}), \quad w_{ij}=\frac{\tilde{w}_{ij}}{\sum_{j'\in\mathcal{N}(i)}\tilde{w}_{ij'}},
\label{eq:concept_mod_weight}
\end{equation}
where the hyperparameter $\lambda$ controls the intensity of semantic integration, followed by a normalization step to ensure the total weight for each neighbor set sums to unity. By reinforcing semantically consistent connections, the model constrains propagation within coherent conceptual boundaries.

By designating high-quality representatives as landmarks, we initialize a label matrix $\mathbf{Y} \in \mathbb{R}^{N \times K}$ by setting $Y_{ik} = 1$ if sample $i$ is a representative for concept $k$, and $0$ otherwise.  We perform iterative propagation on the concept-aware graph, employing random walk \cite{zhu-etal-2003-Learning} that treats these initial references as stationary sources of trusted evidence:
\begin{equation}
\mathbf{Q}^{(t+1)}_{i,:} = 
\begin{cases} 
\mathbf{Y}_{i,:}, & i \in \mathcal{S}, \\
\alpha \sum_{j \in \mathcal{N}(i)} w_{ij} \mathbf{Q}^{(t)}_{j,:}, & i \notin \mathcal{S},
\end{cases}
\label{eq:lp_update_piecewise}
\end{equation}
where $\alpha \in (0, 1)$ is a diffusion factor that controls the spatial extent of semantic propagation. In each iteration, every non-anchor sample updates its conceptual belief by aggregating the distributions of its neighbors, weighted by the semantically-refined edges $w_{ij}$. This process facilitates a global diffusion of semantic information across the manifold, which actively guides the model to produce a final posterior $\mathbf{Q}$ that achieves a harmonious integration of high-level conceptual priors and local structural consistency, effectively bridging the gap between global semantics and local topology.

To fully exploit these propagated signals, we employ concept-supervised contrastive learning to regularize the latent space, leveraging the posteriors to simultaneously enforce intra-concept compactness and inter-concept separation. This objective is prioritized on a high-confidence set $\mathcal{H}$ to ensure the semantic purity of the supervision, where $\mathcal{H}$ is formed by retaining the top $\rho$ fraction of instances ranked by their predictive certainty, $\mathrm{conf}_i = \max_k Q_{ik}$. Following the masking strategy established during pre-training, we construct three distinct views $\{\mathbf{z}_i^{TAV}, \mathbf{z}_i^{TA0}, \mathbf{z}_i^{T0V}\}$ for each $i \in \mathcal{H}$ by selectively omitting modalities. This multi-view augmentation method yields robust positive samples that remain invariant to partial modality absence while ensuring the semantic remains consistently grounded by the textual modality.

We treat each propagated assignment $\hat{y}_i = \arg\max_k Q_{ik}$ as a supervisory signal to structure the contrastive embedding space. Let $\mathcal{I}$ denote the set of all augmented views within a minibatch, and $g(\cdot)$ represent a projection head with $\ell_2$ normalization. Defining the positive set for a view $u$ based on shared conceptual assignments as $\mathcal{P}(u) = \{v \in \mathcal{I} \setminus \{u\} \mid \hat{y}(v) = \hat{y}(u)\}$, we optimize the following concept-supervised contrastive objectives:
\begin{equation}
Z_c(u) = \sum_{w\in\mathcal{I}\setminus\{u\}} \exp\left( \mathrm{sim}(g(u), g(w)) / \tau \right),
\label{eq:con_partition}
\end{equation}
\begin{equation}
\ell(u,v) = -\log \left( \frac{\exp \left( \mathrm{sim}(g(u), g(v)) / \tau \right)}{Z_c(u)} \right),
\label{eq:con_pair}
\end{equation}
\begin{equation}
\mathcal{L}_{\mathrm{c\text{-}sup}} = \frac{1}{|\mathcal{I}|} \sum_{u\in\mathcal{I}} \frac{1}{|\mathcal{P}(u)|} \sum_{v\in\mathcal{P}(u)} \ell(u,v).
\label{eq:concept_supcon}
\end{equation}
Minimizing Eq.~\eqref{eq:concept_supcon} transforms the propagated posteriors into a geometric constraint, anchoring the embedding space within a structured conceptual manifold.
By grouping views under the same concept and separating those from competing categories, this optimization creates a feedback loop. Consequently, the manifold aligns with the semantic graph, enabling robust landmark identification and reliable semantic propagation in later iterations.


\section{Experiments}


\subsection{Datasets and Evaluation Metrics}
We evaluate MCSP on three multimodal intent discovery benchmarks: MIntRec, MIntRec2.0, and MELD-DA. As no validation set is used in unsupervised setting, we merge the original train, validation, and test splits and repartition the data 4:1 for training and testing to maximize unlabeled data utilization.
Following \cite{fahad-etal-2014-survey, saxena-etal-2017review}, we employ four standard metrics to assess clustering performance: ACC, ARI, NMI, and FMI.  Detailed dataset statistics and metric definitions
are provided in Appendices~\ref{app:dataset} and~\ref{app:evaluation}.
\begin{table*}[t]
  \caption{Results on three benchmarks. The two MCSP variants differ only in the MLLM used for concept generation, namely Qwen3-VL and Gemini-3.0-Pro. Best results are in \textbf{bold} and second-best are \underline{underlined}.}
  \label{tab:main_results}
  \centering
  \setlength{\tabcolsep}{3.2pt}
  \renewcommand{\arraystretch}{1.05}
  \resizebox{0.97\textwidth}{!}{%
    \begin{tabular}{l|ccccc|ccccc|ccccc}
      \toprule
      Datasets &
      \multicolumn{5}{c|}{MIntRec} &
      \multicolumn{5}{c|}{MIntRec2.0} &
      \multicolumn{5}{c}{MELD-DA} \\
      \midrule
      Methods
      & ACC & ARI & NMI & FMI & Avg.
      & ACC & ARI & NMI & FMI & Avg.
      & ACC & ARI & NMI & FMI & Avg. \\
      \midrule
      SCCL
      & 36.36 & 16.94 & 43.98 & 22.67 & 29.99
      & 21.03 & 8.88 & 30.57 & 13.25 & 18.43
      & 25.79 & 12.23 & 19.10 & 23.93 & 20.26 \\
      CC
      & 40.36 & 21.10 & 46.50 & 25.99 & 33.49
      & 26.23 & 12.51 & 35.55 & 16.37 & 22.67
      & 24.54 & 12.60 & \textbf{23.19} & 23.92 & 21.06 \\
      MCN
      & 17.30 & 2.01 & 18.85 & 9.33 & 11.84
      & 11.47 & 2.32 & 13.63 & 6.99 & 8.60
      & 18.10 & 1.57 & 8.34 & 15.31 & 10.83 \\
      USNID
      & 40.36 & 19.88 & 46.77 & 25.01 & 33.01
      & 21.73 & 9.93 & 32.14 & 14.02 & 19.50
      & 24.40 & 12.17 & 20.18 & 25.01 & 20.44 \\
      SPILL
      & 36.31 & 15.44 & 38.83 & 20.94 & 27.88
      & 22.10 & 7.91 & 26.63 & 12.19 & 17.21
      & 23.11 & 3.97 & 12.97 & 18.02 & 14.52 \\
      UMC
      & 42.92 & 23.35 & 48.17 & 28.11 & 35.64
      & 27.36 & 13.50 & 36.78 & 17.34 & 23.75
      & 33.79 & 18.45 & 21.38 & 31.80 & 26.36 \\
      \midrule
      MCSP$_{\text{Qwen}}$
      & \underline{44.54} & \underline{24.85} & \textbf{49.07} & \underline{29.61} & \underline{37.02}
      & \textbf{29.54} & \underline{15.00} & \textbf{37.37} & \textbf{18.96} & \textbf{25.22}
      & \textbf{34.62} & \textbf{22.07} & 21.49 & \textbf{33.74} & \textbf{27.98} \\
      MCSP$_{\text{Gemini}}$
      & \textbf{45.21} & \textbf{25.59} & \underline{48.42} & \textbf{30.28} & \textbf{37.37}
      & \underline{29.19} & \textbf{15.03} & \underline{37.33} & \underline{18.92} & \underline{25.12}
      & \underline{34.57} & \underline{21.57} & \underline{21.61} & \underline{33.16} & \underline{27.73} \\
      \bottomrule
    \end{tabular}%
  }
\end{table*}

\begin{table*}[t]
  \caption{Ablation results on three benchmark datasets. Best results are in \textbf{bold} and second-best are \underline{underlined}.}
  \label{tab:ablation_three_datasets}
  \centering
  \setlength{\tabcolsep}{2.6pt}
  \renewcommand{\arraystretch}{1.05}
  \resizebox{0.97\textwidth}{!}{%
    \begin{tabular}{l|ccccc|ccccc|ccccc}
      \hline
      Datasets & \multicolumn{5}{c|}{MIntRec} & \multicolumn{5}{c|}{MIntRec2.0} & \multicolumn{5}{c}{MELD-DA} \\
      \hline
      Ablation & ACC & ARI & NMI & FMI & Avg. & ACC & ARI & NMI & FMI & Avg. & ACC & ARI & NMI & FMI & Avg. \\
      \hline
      w/o Step1
      & 17.89 & 2.31  & 20.84 & 9.83  & 12.72
      & 9.97  & 0.79  & 10.52 & 6.07  & 6.84
      & 19.64 & 0.58  & 6.03  & 17.18 & 10.86 \\
      w/o $\mathcal{T}_{\text{refined}}$
      & 41.03 & 21.50 & 45.53 & 26.46 & 33.63
      & 28.41 & 13.58 & 36.47 & 17.48 & 23.99
      & \underline{31.68} & \underline{17.56} & 20.83 & \underline{29.64} & \underline{24.93} \\
      w/o $\lambda$
      & \underline{43.46} & \underline{23.55} & \underline{46.97} & \underline{28.35} & \underline{35.58}
      & \underline{28.56} & \underline{13.94} & \underline{36.88} & \underline{17.83} & \underline{24.30}
      & 30.50 & 16.81 & \underline{20.86} & 28.68 & 24.21 \\
      w/o $\mathcal{G}$
      & 24.27 & 7.85  & 30.16 & 16.08 & 19.59
      & 17.86 & 5.37  & 22.18 & 12.25 & 14.42
      & 28.50 & 5.36  & 11.20 & 27.70 & 18.19 \\
      w/o $\mathcal{H}$
      & 40.90 & 21.74 & 44.66 & 26.67 & 33.49
      & 26.27 & 11.85 & 34.05 & 15.88 & 22.01
      & 30.97 & 15.53 & 19.00 & 28.06 & 23.29 \\
      \hline
      MCSP$_{\text{Gemini}}$
      & \textbf{45.21} & \textbf{25.59} & \textbf{48.42} & \textbf{30.28} & \textbf{37.37}
      & \textbf{29.19} & \textbf{15.03} & \textbf{37.33} & \textbf{18.92} & \textbf{25.12}
      & \textbf{34.57} & \textbf{21.57} & \textbf{21.61} & \textbf{33.16} & \textbf{27.73} \\
      \hline
    \end{tabular}%
  }
\end{table*}

\subsection{Baselines}
We compare our method against state-of-the-art unsupervised clustering methods, including: 
(1) SCCL \citep{zhang-etal-2021-supporting} integrates top down clustering with instance contrastive learning to enhance cluster separation by tightening local neighborhoods.
(2) CC \citep{li-etal-2021-contrastive} optimizes contrastive objectives for both instances and clusters on augmented views to achieve dual alignment.
(3) MCN \citep{chen-etal-2021-multimodal} learns a unified multimodal embedding through cross modal projection and explicit clustering to group semantically related instances.
(4) USNID \citep{zhang-etal-2024-clustering} alternates between centroid guided grouping and self supervised learning to generate consistent pseudo labels following contrastive pretraining.
(5) SPILL \citep{lin-etal-2025-spill} uses an LLM to select utterances with the same intent as the target from a set of retrieved candidates, and pools them with the target utterance to improve representations for intent clustering without additional fine-tuning.
(6) UMC \citep{zhang-etal-2024-unsupervised} defines the current performance frontier by combining contrastive pre-training and a density driven curriculum to progressively optimize multimodal clustering representations.


\subsection{Experimental Settings}

To improve reproducibility and reduce the need to tune parameters separately for each dataset in the unsupervised setting, we keep most hyperparameters the same across datasets and adjust only a few parameters that are related to training stability. In the concept generation stage, we use Gemini-3.0-Pro and Qwen-3-VL for semantic reasoning, with the candidate size $m$ fixed at 40 for all datasets. To reduce cost, we retain only three representatives for semantic discovery. An analysis of time efficiency and resource consumption is provided in Appendix \ref{cost}. For semantic propagation, we construct a $k$-NN graph with $k=10$ and set $\lambda=3.0$, while uniformly fixing $\alpha=0.95$ and propagation steps at 40 across all datasets.

For MIntRec, MIntRec2.0, and MELD-DA, the pre-training temperatures are set to (0.2, 0.2, 0.3) with corresponding learning rates of ($5\mathrm{e}{-5}$, $2\mathrm{e}{-5}$, $5\mathrm{e}{-5}$). All experiments employ a batch size of 128 and utilize the AdamW optimizer \cite{loshchilov-etal-2019-decoupled} for 12 epochs without gradient clipping. The training learning rates are ($1\mathrm{e}{-4}$, $1\mathrm{e}{-4}$, $5\mathrm{e}{-5}$), and the $\rho$ is fixed at 0.6 with its sensitivity analysis detailed in Appendix D.2. The temperature triplets $(\tau, \tau_c, \tau_g)$ are configured as (0.5, 1.0, 0.7), (0.3, 0.3, 7.5), and (2.5, 1.0, 3.5), respectively, as discussed in Section~\ref{sec:sensitivity}. All experiments are conducted on an NVIDIA Tesla V100 GPU, and results are averaged over five runs with seeds from 0 to 4, with a stability analysis provided in Appendix \ref{MCSPSTA}.

\subsection{Results}

Table~\ref{tab:main_results} shows that both MCSP variants consistently
outperform UMC, the strongest baseline. Specifically,
MCSP$_{\text{Qwen}}$ and MCSP$_{\text{Gemini}}$ surpass UMC by 1.49\% and 1.48\%, respectively. Their
comparable performance across various MLLMs indicates that MCSP
is robust to the choice of reasoning model. Compared with SPILL, which also leverages LLM-based semantics, both variants improve the average score by $10.20\%$, highlighting the effectiveness of MCSP's semantic
guidance.
Notably, MCSP also provides better interpretability by aligning clusters with explicit concepts, as discussed in Section~\ref{app:case_study}.

On MIntRec, MCSP$_{\text{Gemini}}$ achieves a substantial performance breakthrough in ACC, outperforming UMC by 2.29\%. This improvement indicates that the semantic anchors derived from MLLMs provide superior semantic grounding, which effectively aligns clusters with their underlying intent categories. Furthermore, MCSP$_{\text{Qwen}}$  yields a peak NMI of 49.07, underscoring its capacity to capture intricate mutual information within the multimodal feature space.
The efficacy of our approach is further evidenced on the more complex MIntRec2.0 by a significant leap in ARI. Specifically, MCSP$_{\text{Qwen}}$ reaches an ARI of 15.00, surpassing UMC by a 1.50\% margin. Since ARI serves as a stringent measure of cluster structural integrity, this gain demonstrates that our high-level semantic propagation effectively reconstructs a more robust manifold structure, ensuring coherent intent discovery in high-dimensional scenarios with increased label complexity.
On MELD-DA, MCSP$_{\text{Qwen}}$ again achieves the best average result, with gains of 3.62\% in ARI, 0.83\% in ACC, and 1.94\% in FMI over UMC, indicating improved cluster purity under ambiguous conversational labels. Conversely, the slight NMI deficit relative to CC stems from the coarse granularity of MELD-DA, which limits the reasoning precision during concept generation.


\section{Discussion}
\label{sec:discussion}

\subsection{Ablation Studies}
\label{sec: abalition}

Table~\ref{tab:ablation_three_datasets} summarizes the individual contributions of each component within the MCSP$_{\text{Gemini}}$.

\textbf{w/o Step1.} 
The most severe performance collapse occurs upon removing the pre-training with average scores plummeting to 10.86\% on MELD-DA and 6.84\% on MIntRec2.0. Such substantial drops including an average ARI decrease exceeding 19\% underscore the indispensable role of a cluster-friendly backbone.


\textbf{w/o $\mathcal{T}_{\text{refined}}$.} 
Substituting semantic concepts with dummy placeholders, such as "Concept 0" or "Concept 1", leads to a sharp performance drop, especially on MIntRec where ACC and ARI decrease by over 4\%. This gap confirms the necessity of generating high quality concepts rather than relying on arbitrary labels. To further address concerns regarding concept consistency, we provide a detailed stability analysis in Appendix \ref{concept_sta}.

\textbf{w/o $\lambda$.} 
Removing semantic edge modulation degrades performance, particularly on MELD-DA where metrics suffer a sharp decline exceeding 4\%. This degradation confirms that semantic modulation is essential to align neighborhood geometry with conceptual consistency for more reliable propagation.

\textbf{w/o $\mathcal{G}$.} \label{method:graph_ablation}
Replacing graph propagation with direct nearest assignment triggers a substantial performance collapse including average scores plummeting by 17.78\% 10.70\% and 9.54\% on three datasets. This sharp decline underscores the value of semantic propagation for expanding concepts into globally consistent intent clusters.

\textbf{w/o $\mathcal{H}$.} 
Removing confidence filtering reduces average performance by approximately 4\%. Notably, the ARI on MELD-DA plummets by 6\%, underscoring the necessity of filtering to prevent error reinforcement and maintain discriminative representations.



\begin{figure*}[t]
  \centering
  \includegraphics[width=\textwidth]{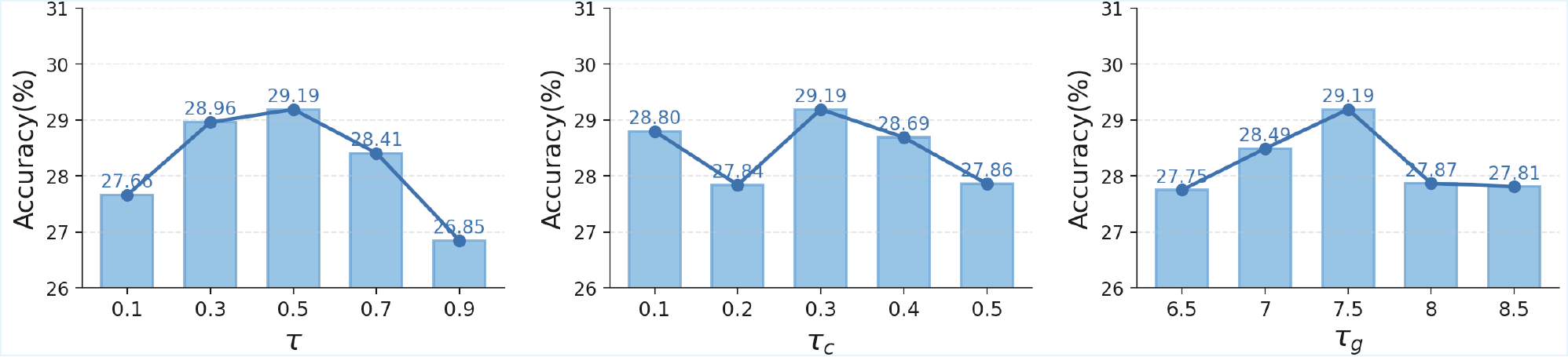}
  \caption{Temperature hyper-parameter sensitivity analysis on MIntRec2.0 measured by accuracy (\%).}
  \Description{A sensitivity analysis plot showing the effect of the temperature hyper-parameter on accuracy on the MIntRec2.0 dataset.}
  \label{fig:sens_temperatures}
\end{figure*}

\begin{figure*}[t]
  \centering
  \includegraphics[width=\textwidth]{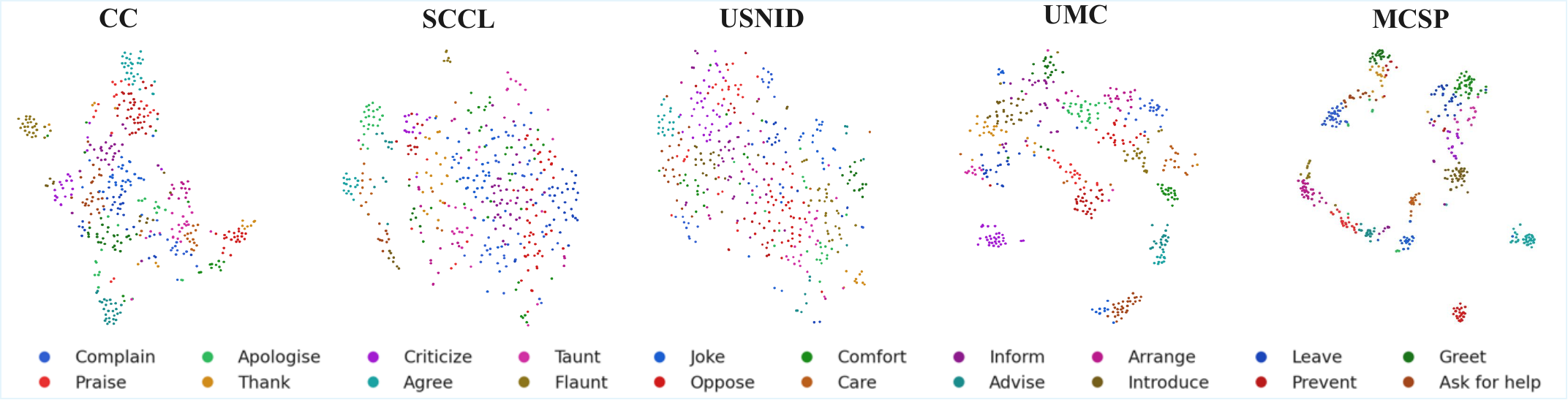}
  \caption{Visualization of representations on the MIntRec dataset.}
  \Description{A visualization of learned representations on the MIntRec dataset, illustrating the distribution and separation of different intent clusters in the embedding space.}
  \label{fig:tsne_mintrec}
\end{figure*}

\subsection{Sensitivity Analysis}
\label{sec:sensitivity}
We investigate the sensitivity of MCSP to three temperature
hyperparameters $\tau$, $\tau_c$, and $\tau_g$, which control
distribution smoothness in contrastive learning, concept assignment,
and graph propagation, respectively. Figure~\ref{fig:sens_temperatures}
reports their effects on clustering performance, while
Appendix~\ref{app:temperature_selection} provides a complementary
label-free selection analysis based on clustering stability across
random seeds.

Specifically, for the $\tau$ in concept-supervised contrastive learning, the accuracy reaches a peak of 29.19\% when $\tau$ is set to 0.5. Smaller values such as 0.1 or 0.3 create overly sharp distributions that penalize negative pairs too aggressively, while larger values like 0.7 or 0.9 lead to excessively smooth distributions that fail to capture fine-grained semantic differences. Similarly, the $\tau_c$ achieves its optimal performance of 29.19\% at 0.3, whereas deviating to 0.2 or 0.4 results in performance drops to 27.84\% and 28.69\% respectively. This indicates the necessity of a precisely calibrated confidence level when associating samples with latent concepts. Regarding the $\tau_g$ in semantic propagation, the model shows a clear preference for a setting of 7.5 to achieve 29.19\% accuracy. Values such as 6.5 or 8.5 lead to suboptimal pseudo-label generation by either restricting information diffusion or allowing non-discriminative signals to propagate through the graph.
These results demonstrate that while MCSP maintains robustness, precise temperature calibration remains vital to balance discriminative power and semantic consistency in unsupervised discovery.

\subsection{Visualization}

Figure~\ref{fig:tsne_mintrec} visualizes method embeddings via t-SNE
\cite{van_der_maaten-2008-visualizing} on MIntRec. While CC and SCCL
exhibit heavily entangled clusters where semantic boundaries are almost
indistinguishable, USNID and UMC improve intra-class coherence but still
suffer from interleaved margins and scattered outliers. In contrast,
MCSP$_{\text{Gemini}}$ generates the most structured geometry,
characterized by compact regions and significantly reduced overlap
across the latent space.


Specifically, MCSP$_{\text{Gemini}}$ successfully isolates distinct intents such as Praise, Greet, and Inform into compact and separated clusters. This outcome confirms that semantic propagation mechanism anchors clear manifolds by spreading MLLM generated conceptual priors throughout the latent space. Although minor overlap persists between semantically similar categories like Complain, Criticize, and Taunt, the overall spatial structure remains highly organized. These visualizations mirror the quantitative gains shown in Table~\ref{tab:main_results}, demonstrating that constrained semantic flow effectively reshapes the embedding space into coherent intent manifolds.
Visualizations on MIntRec2.0 and MELD-DA datasets are provided in Appendix \ref{APPD}.

\subsection{Case Study}
\label{app:case_study}

\begin{figure*}[t]
  \centering
  \includegraphics[width=\textwidth]{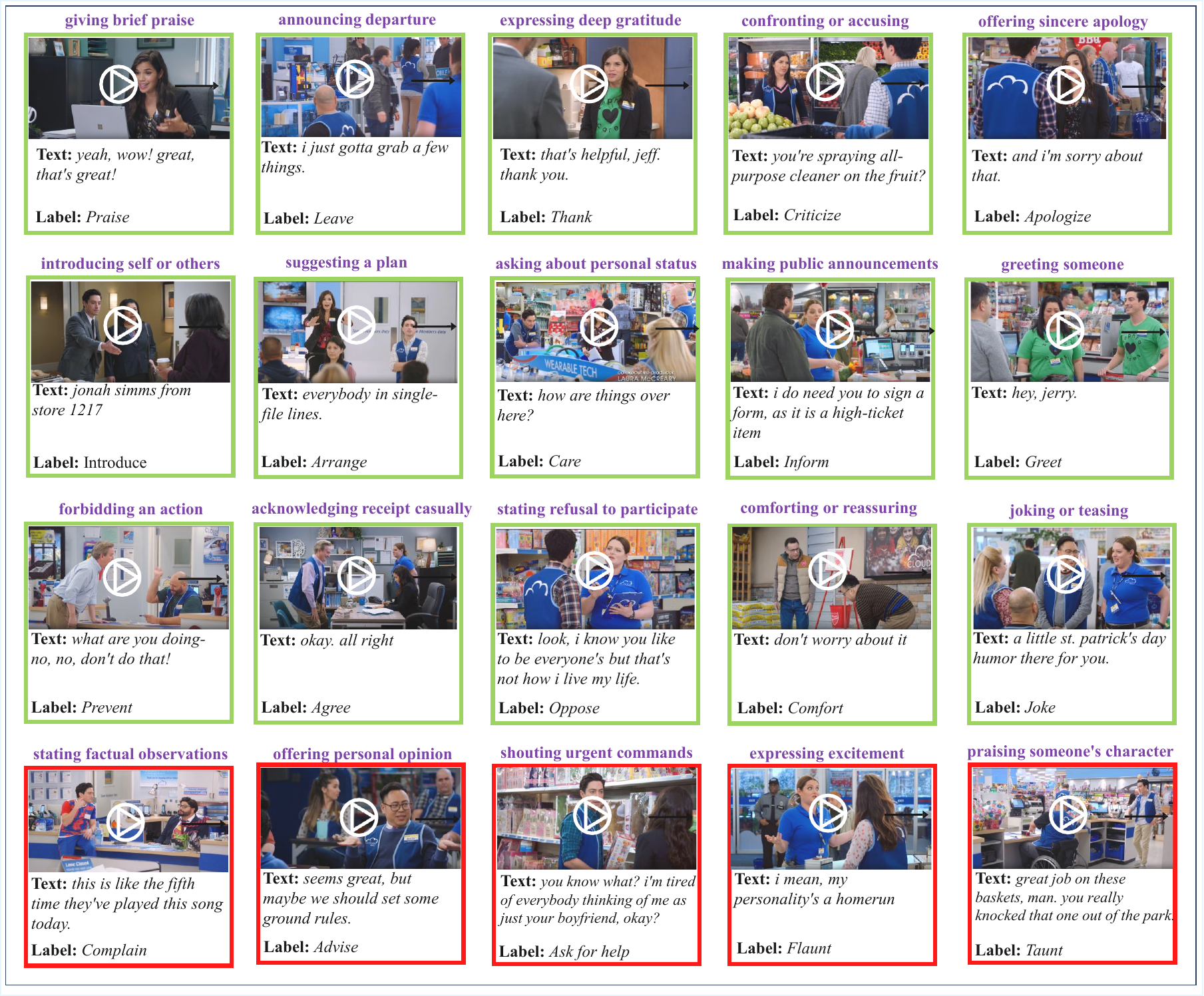}
  \caption{Case study on MIntRec. Representative samples display video, text, and GT labels. Purple text indicates generated concepts. Green and red borders indicate strong alignment and semantic deviation from GT labels, respectively.}

  \label{fig:case_study}
\end{figure*}

As illustrated in Figure~\ref{fig:case_study},
we conduct a qualitative case study on the MIntRec dataset to evaluate the interpretability and semantic faithfulness of the concepts generated by MCSP.

The empirical evidence suggests that MCSP produces robust concept anchors that are consistent with ground truth labels and deeply grounded in multimodal context. A compelling example is the \textbf{Leave} intent, where brief and semantically underspecified utterances are correctly identified as departure related acts by integrating visual cues that provide decisive evidence when lexical content is insufficient. More importantly, the contrastive reasoning mechanism effectively disambiguates closely related intents by isolating unique functional traits. For instance, in the \textbf{Criticize} cluster, while the text "spraying cleaner on fruit" might be misinterpreted as a neutral description, our model contrasts it against neighboring clusters to generate the precise functional label "confronting or accusing." Similarly, for \textbf{Introduce}, the model leverages multimodal alignment to correctly abstract "introducing self or others," demonstrating that our representative mining yields stable anchors for context aware interpretation. In the challenging \textbf{Joke} category, MCSP successfully captures underlying humor despite its proximity to teasing, while the \textbf{Apologize} intent is distinguished by its sincere tone from more casual acknowledgments. These successful cases validate that producing explicit, human readable concept anchors through contrastive inference significantly enhances both the discriminative power and transparency of discovered clusters.

Despite these strengths, certain failure modes reveal inherent challenges in unsupervised semantic discovery. First, the model occasionally struggles to decode deep pragmatic layers such as sarcasm or social stance, potentially misinterpreting \textbf{Taunt} as praise or reducing \textbf{Flaunt} to generic excitement due to an over-reliance on surface-level affect. Second, without a predefined label inventory, fine-grained intent boundaries are difficult to align, leading the model to summarize \textbf{Advise} as a general opinion rather than a specific directive act. Third, the quality of concept generation remains sensitive to cluster noise and class imbalance, where non-prototypical representatives can bias the output, such as misidentifying \textbf{Complain} as a neutral factual observation. These constraints are quite helpful, as they effectively highlight specific areas for refining unsupervised clustering and concept alignment.

\section{Conclusion}
In this work, we propose MCSP, a novel method for multimodal intent discovery that addresses limited interpretability and over-reliance on geometry-driven refinement by integrating MLLM-Guided concept generation with semantic propagation. First, MCSP identifies high-quality representatives through a progressive refinement strategy to provide a stable foundation for MLLM-guided contrastive reasoning, yielding explicit concepts that serve as global semantic anchors. Second, these conceptual priors are diffused across the data manifold via a semantic propagation mechanism, refining neighborhood geometry through semantic alignment to harmonize global guidance with local structural consistency. Integrated into this process, a synergistic feedback loop between random walk based diffusion and concept-supervised contrastive learning continuously regularizes the latent space, progressively shaping the embedding manifold for a reliable and transparent identification of intents. Extensive benchmarks validate that MCSP consistently outperforms state-of-the-art methods while generating semantically rich concepts that provide intuitive justifications for the discovered clusters, thereby establishing a pioneering and interpretable paradigm for multimodal discovery.

\bibliographystyle{ACM-Reference-Format}
\bibliography{custom}

\appendix

\section{Appendix Overview}
\label{sec:appendix_overview}


This appendix provides supplementary materials that strengthen the
methodological and empirical foundations of MCSP. It extends the
theoretical background, documents implementation and evaluation
protocols, justifies key design choices, and presents
analyses of representation quality, robustness, and computational
efficiency.

Section~\ref{sec:appendix} reviews related works on open-world intent
discovery and LLM-guided clustering. Section~\ref{APPE} documents the
prompt design, benchmark statistics and clustering
metrics used in our experiments. Section~\ref{APP:C} analyzes key design
choices, including sample quality scoring, confidence-based filtering,
and label-free temperature selection. Section~\ref{APPD} presents t-SNE
visualizations on MIntRec2.0 and MELD-DA. Section~\ref{APP} evaluates
robustness under an estimated category number and analyzes the stability
of clustering results and generated concepts across random seeds.
Finally, Section~\ref{cost} reports the runtime and resource consumption
of MCSP.


\section{Additional Related Works
}
\label{sec:appendix}

To augment the foundational unsupervised methodologies delineated in Section \ref{related} and construct a more comprehensive theoretical landscape, we extend our review to encompass the specialized domains of open-world intent discovery and LLM-augmented clustering, thereby providing a holistic contextual background.

\subsection{Open-world Intent Discovery}
\label{app:openworld_intent}

Beyond pure clustering, deployed dialogue systems often operate in an open-world regime where in-domain intents coexist with emerging out-of-domain queries. Early studies primarily evaluated a model's ability to detect out-of-scope inputs at inference time \citep{larson-etal-2019-evaluation}. In the broader open intent recognition literature, the problem is organized as a two-stage pipeline: open intent detection separates known and unknown instances, and open intent discovery further groups the unknown instances into coherent categories. TEXTOIR provides an integrated and visualized platform that unifies these two modules and supports an end-to-end pipeline implementation \citep{zhang-etal-2021-textoir}.
Within this family of tasks, New Intent Discovery (NID) typically assumes labeled in-domain intents to facilitate clustering over unlabeled data, while Generalized Intent Discovery (GID) further requires simultaneously recognizing labeled in-domain intents and discovering new out-of-domain types under a unified open-world intent set \citep{mou-etal-2022-generalized}. To better match streaming deployment, Continual Generalized Intent Discovery extends GID to dynamic stages, requiring incremental integration of newly emerging intents with minimal access to historical data \citep{song-etal-2023-continual}.

\subsection{LLM-Guided Clustering}
\label{app:llm_clustering}

Recent progress in large language models has inspired clustering methods that incorporate external semantic knowledge beyond purely embedding-based pipelines. ClusterLLM leverages instruction-tuned LLMs to provide pairwise and triplet-style guidance for refining cluster assignments and selecting appropriate clustering granularity \citep{zhang-etal-2023-clusterllm}. Complementary work studies query-efficient or few-shot settings, where limited demonstrations or supervision are amplified through LLM-generated guidance to steer clustering with reduced annotation effort \citep{viswanathan-etal-2024-large}. For hierarchical clustering, Pattnaik et al.\ introduce LLM-guided multi-view cluster representations to improve the quality of hierarchical text clustering \citep{pattnaik-etal-2024-improving}.
Beyond direct assignment refinement, LLMs have also been used to enhance cluster structure at difficult regions, such as refining ambiguous boundary instances to stabilize clustering decisions \citep{feng-etal-2024-llmedgerefine}. Related ideas have been explored in adjacent settings such as topic modeling, where LLM guidance is used to inject semantics into the clustering process for more coherent topic discovery \citep{liu-etal-2025-llm-guided}.

While LLM-guided clustering has shown promise in text only scenarios through local constraints or post hoc refinement, its application to multimodal intent discovery remains largely under explored and often faces prohibitive query costs. MCSP addresses this gap by shifting from discrete local signals to propagatable cluster semantics derived from a sparse set of representatives. By disseminating these concepts through a structured propagation mechanism, our method establishes a globally coherent semantic structure integrated with representation learning, providing a scalable solution for aligning complex tri modal signals.


\section{Implementation and Evaluation Details}
\label{APPE}
This section documents the prompt design and evaluation protocol of
MCSP, including the multimodal reasoning prompts, benchmark statistics and clustering evaluation metrics.

\subsection{Details of Prompt Design}
\label{sec:prompt_design}

\begin{figure}[t]
\centering
\begin{tcolorbox}[promptbox={Prompt for Multimodal Contrastive Reasoning}]
\footnotesize
\raggedright
\setlength{\parindent}{0pt}

\textbf{Instruction.}
You are an expert in \textbf{UNSUPERVISED SEMANTIC DISCOVERY} for multimodal dialogues. You will be given short videos and their spoken text for one \textbf{FOCUS CLUSTER} (Group A) and one \textbf{NEAREST NEIGHBOR CLUSTER} (Group B).

\textbf{Goal.}
Extract \textbf{ONE ABSTRACT} communicative-function label for Group A ONLY. Group B is a contrast set provided ONLY to improve the discriminativeness of the label for Group A.

\textbf{Evidence \& Alignment (Video + Spoken Text).}
\begin{itemize}
  \item Videos are provided in order: \texttt{\{video\_order\_str\}}. Match them to tags (A1/A2/B1).
  \item If video and spoken text conflict, prioritize the video.
\end{itemize}

\textbf{Definition.}
The semantic is the \emph{communicative function}: the speaker's pragmatic goal or speech act. It should be at the \textbf{dialogue-act} level, not topic, story, or concrete situation.

\textbf{Constraints.}
\begin{itemize}
  \item Do not describe scenarios, locations, or concrete events (no specific entities).
  \item Do not describe topic content or summarize the story.
  \item Do not mention dataset label names, codes, or taxonomies.
  \item Do not write a full sentence or mix multiple functions.
\end{itemize}

\textbf{Style.}
Output a \textbf{2--6 word} English phrase using a \textbf{canonical gerund form} (e.g., "requesting help"). It must be generic, reusable, and discriminative.

\textbf{Inputs.}
\begin{tabularx}{\linewidth}{@{}l X@{}}
\textbf{Group A (Focus):} & ID \texttt{\{cid\}}; Spoken text: \texttt{\{A\_block\}} \\
\textbf{Group B (Neighbor):} & ID \texttt{\{nb\}}; Spoken text: \texttt{\{B\_block\}} (Distance: \texttt{\{d\}}) \\
\end{tabularx}

\textbf{Task.}
(1) Infer the shared function in Group A; (2) compare against Group B and \textbf{remove} any function also applicable to B; (3) output one unique fine-grained label for A.

\textbf{Output (strict JSON).}
\begin{tcolorbox}[codebox]
\begin{Verbatim}[fontsize=\scriptsize,baselinestretch=1]
[
  {
    "cid": {cid},
    "intent": "<2-6 word verb+ing phrase>"
  }
]
\end{Verbatim}
\end{tcolorbox}
\end{tcolorbox}
\caption{$\text{Prompt}_{\text{cont}}$ for multimodal contrastive reasoning, focusing on cluster A against nearest neighbor B.}
\label{fig:prompt_multimodal_contrast}
\end{figure}

\begin{figure}[t]
\centering
\begin{tcolorbox}[promptbox={Prompt for Global Format Normalization}]
\footnotesize
\raggedright
\setlength{\parindent}{0pt}

\textbf{Instruction.}
You are an expert in label processing. Your job is \textbf{FORMAT NORMALIZATION ONLY}. This is \textbf{NOT} semantic merging, \textbf{NOT} re-labeling, and \textbf{NOT} re-clustering.

\textbf{Mandatory Edits (Format-Only).}
\begin{itemize}
  \item Lowercase the label; remove punctuation, quotes, and articles (a/an/the).
  \item Enforce \textbf{2--6 English words} in a \textbf{verb phrase} or \textbf{gerund-style} phrase.
  \item Remove extra filler words that do not contribute to the core intent.
\end{itemize}

\textbf{Negative Constraints.}
\begin{itemize}
  \item \textbf{Do NOT} change the underlying meaning or replace words with synonyms.
  \item \textbf{Do NOT} generalize/broaden labels or add new semantic content.
  \item \textbf{Do NOT} make different \texttt{cid}s share the same label unless they were already identical.
  \item \textbf{Do NOT} add, drop, or modify any \texttt{cid} from the input.
\end{itemize}

\textbf{Inputs.}
\begin{tabularx}{\linewidth}{@{}l X@{}}
\textbf{Input JSON:} & \texttt{\{json\_items\_block\}} \\
\end{tabularx}

\textbf{Output (strict JSON).}
\begin{tcolorbox}[codebox]
\begin{Verbatim}[fontsize=\scriptsize,baselinestretch=1]
[
  {
    "cid": <int>,
    "refined_intent": "<normalized label>"
  },
  ...
]
\end{Verbatim}
\end{tcolorbox}
\end{tcolorbox}
\caption{$\text{Prompt}_{\text{global}}$ used to ensure cross-cluster label consistency and syntactical alignment.}
\label{fig:prompt_normalization}
\end{figure}

We propose a two stage prompting strategy to generate cluster level concepts that are both semantically precise and operationally robust in an unsupervised setting. The first stage, Semantic Discovery, leverages a multimodal large language model to infer communicative functions through contrastive reasoning between a target cluster and its nearest neighbor. The second stage, Stylistic Normalization, ensures output stability by standardizing surface forms without compromising the original meaning. Together, this integrated pipeline transforms raw cluster representatives into a compact, grounded, and discriminative concept inventory.

In the semantic discovery stage, we explicitly define target semantics as dialogue act level functions, steering the model away from topical summaries toward pragmatic intent. To ensure cluster wide generalization, we impose strict exclusion rules against sample specific entities while enforcing multimodal grounding through an alignment protocol that binds video segments to transcripts. Crucially, a contrastive mechanism tasks the model with isolating unique functions and subtracting shared traits from nearest neighbors, thereby avoiding generic outputs in favor of discriminative labels. The output is constrained to a 2 to 6 word gerund phrase in strict JSON format to ensure deterministic parsing. Subsequently, the normalization stage performs format only refinements, such as standardizing casing and length, while strictly prohibiting synonym replacement to prevent semantic drift. By integrating deep semantic inference with systematic normalization, our strategy achieves a synergistic effect that enhances the interpretability and discriminativeness of the unsupervised discovery process.

\subsection{Dataset Details}
\label{app:dataset}

Table~\ref{tab:dataset_statistics} summarizes the statistics of the
three benchmark datasets. MIntRec~\cite{zhang-etal-2022-mintrec} is a
fine-grained multimodal intent dataset with 20 categories. MIntRec2.0
\cite{zhang-etal-2024-mintrec2} extends this benchmark to 30 intent
categories. MELD-DA~\cite{Saha-etal-2020-towards} is a multimodal
dialogue-act dataset derived from the Switchboard corpus
\cite{godfrey-etal-1992-SWITCHBOARD} and contains 12 categories.
For all datasets, we merge the original training and validation splits
for unlabeled training and retain the original test split for
evaluation. No ground-truth labels are used during
training or clustering.

\subsection{Evaluation Details}
\label{app:evaluation}

We evaluate clustering performance using four standard metrics: Accuracy (ACC), Normalized Mutual Information (NMI), Adjusted Rand Index (ARI), and Fowlkes-Mallows Index (FMI). For all metrics, higher values indicate superior quality.

\textbf{Accuracy (ACC)} measures the maximum alignment between predicted labels and ground-truth via the Hungarian algorithm \citep{zhang-etal-2021-supporting, zhang-etal-2024-clustering}. It is defined as the proportion of correctly assigned samples under the optimal permutation.

\textbf{Normalized Mutual Information (NMI)} quantifies the shared information between predicted and true label distributions, normalized by their average entropy:
\begin{equation}
\text{NMI}(y_{gt}, y_{p}) = \frac{\text{MI}(y_{gt}, y_{p})}{\frac{1}{2} \left[ H(y_{gt}) + H(y_{p}) \right]},
\end{equation}
where $y_{gt}$ and $y_{p}$ denote ground-truth and predicted labels, respectively, $\text{MI}(\cdot)$ represents mutual information, and $H(\cdot)$ denotes entropy.

\textbf{Adjusted Rand Index (ARI)} evaluates the similarity between two assignments by considering all pairs of samples and adjusting for chance, let $S = \big[ \sum_{i} \binom{u_i}{2} \sum_{j} \binom{v_j}{2} \big] / \binom{n}{2}$:

\begin{equation}
\text{ARI} = \frac{\sum_{i,j} \binom{n_{i,j}}{2} - S}{\frac{1}{2} \left[ \sum_{i} \binom{u_i}{2} + \sum_{j} \binom{v_j}{2} \right] - S},
\end{equation}

where $n_{i,j}$ is the number of common samples between clusters, while $u_i$ and $v_j$ are the marginal totals of the contingency table.

\textbf{Fowlkes-Mallows Index (FMI)} assesses clustering quality by calculating the geometric mean of pairwise precision and recall:
\begin{equation}
\text{FMI} = \frac{TP}{\sqrt{(TP + FP)(TP + FN)}},
\end{equation}
where $TP$, $FP$, and $FN$ represent the number of true positives, false positives, and false negatives, respectively.
\begin{table}[t]
  \centering
  \caption{Statistics of the benchmark datasets. \#C and \#U denote
  the numbers of categories and utterances, respectively.}
  \label{tab:dataset_statistics}
  \small
  \setlength{\tabcolsep}{4pt}
  \begin{tabular}{lrrrr}
    \toprule
    Dataset & \#C & \#U & Train & Test \\
    \midrule
    MIntRec    & 20 & 2,224 & 1,779 & 445 \\
    MIntRec2.0 & 30 & 9,304 & 7,271 & 2,033 \\
    MELD-DA    & 12 & 9,988 & 7,990 & 1,998 \\
    \bottomrule
  \end{tabular}
\end{table}
\begin{table*}[t]
\centering
\caption{Top-18 label-free temperature stability results on MIntRec.}
\label{tab:temperature_stability}
\scriptsize
\setlength{\tabcolsep}{2.5pt}
\renewcommand{\arraystretch}{1.08}
\resizebox{\textwidth}{!}{%
\begin{tabular}{cc cc cc cc cc cc}
\toprule
$(\tau,\tau_c,\tau_g)$ & Stab. &
$(\tau,\tau_c,\tau_g)$ & Stab. &
$(\tau,\tau_c,\tau_g)$ & Stab. &
$(\tau,\tau_c,\tau_g)$ & Stab. &
$(\tau,\tau_c,\tau_g)$ & Stab. &
$(\tau,\tau_c,\tau_g)$ & Stab. \\
\midrule

\textbf{(0.3, 1.5, 1.0)} & \textbf{0.4000} &
(0.3, 1.0, 0.5) & 0.3978 &
(0.3, 0.5, 0.5) & 0.3977 &
(0.3, 1.5, 0.5) & 0.3972 &
(0.5, 0.5, 0.7) & 0.3933 &
(0.5, 1.5, 0.7) & 0.3921 \\

(0.5, 0.5, 0.5) & 0.3914 &
(0.3, 1.0, 1.0) & 0.3908 &
(0.5, 1.0, 0.7) & 0.3906 &
(0.3, 1.0, 0.7) & 0.3905 &
(0.3, 0.5, 1.0) & 0.3880 &
(0.3, 0.5, 0.7) & 0.3865 \\

(0.5, 1.5, 1.0) & 0.3831 &
(0.3, 1.5, 0.7) & 0.3828 &
(0.7, 1.0, 0.5) & 0.3816 &
(0.5, 0.5, 1.0) & 0.3800 &
(0.5, 1.0, 0.5) & 0.3794 &
(0.7, 1.0, 1.0) & 0.3749 \\

\bottomrule
\end{tabular}%
}
\end{table*}

\section{Quality Score and Hyperparameter Analysis}
\label{APP:C}
This section analyzes key design choices and hyperparameters that affect
pseudo-label reliability and optimization stability, including sample
quality scoring, confidence-based filtering, and label-free temperature
selection.

\begin{table*}[t]
  \caption{Comparison of quality scores for sample selection on three benchmark datasets, including density score, central score, and margin score. Best results are in \textbf{bold} and second-best are \underline{underlined}.}
  \label{tab:quality_score_ablation_three_datasets}
  \centering
  \setlength{\tabcolsep}{3.2pt}
  \renewcommand{\arraystretch}{1.05}
  \resizebox{0.97\textwidth}{!}{%
    \begin{tabular}{l|ccccc|ccccc|ccccc}
      \hline
      Datasets & \multicolumn{5}{c|}{MIntRec} & \multicolumn{5}{c|}{MIntRec2.0} & \multicolumn{5}{c}{MELD-DA} \\
      \hline
      Score & ACC & ARI & NMI & FMI & Avg. & ACC & ARI & NMI & FMI & Avg. & ACC & ARI & NMI & FMI & Avg. \\
      \hline
      density
      & \underline{45.08} & \underline{25.19} & \underline{48.21} & \underline{30.07} & \underline{37.14}
      & 26.10 & 12.26 & 33.36 & 16.34 & 22.92
      & 28.49 & 11.12 & 16.83 & 24.38 & 20.20 \\
      central
      & 42.56 & 23.28 & 46.89 & 28.11 & 35.21
      & \underline{28.91} & \underline{14.83} & \underline{36.60} & \underline{18.72} & \underline{24.77}
      & \underline{31.60} & \underline{16.71} & \underline{20.61} & \underline{29.18} & \underline{24.53} \\
      \hline
      margin
      & \textbf{45.21} & \textbf{25.59} & \textbf{48.42} & \textbf{30.28} & \textbf{37.37}
      & \textbf{29.19} & \textbf{15.03} & \textbf{37.33} & \textbf{18.92} & \textbf{25.12}
      & \textbf{34.57} & \textbf{21.57} & \textbf{21.61} & \textbf{33.16} & \textbf{27.73} \\
      \hline
    \end{tabular}%
  }
\end{table*}

\subsection{Quality Scores for Sample Selection}
\label{sec_score}

To mitigate the influence of ambiguous instances in overlapping latent regions, we evaluate sample reliability through three distinct scoring functions. The margin score, as defined in Section \ref{sec:concept_anchors}, characterizes inter-cluster separability by measuring the distance gap between an instance and its closest competing centroids. Alternatively, we consider the central score, which favors prototypical instances by calculating the negative distance from an instance $\mathbf{z}_i$ to its assigned centroid $\boldsymbol{\mu}_{a_i}$ as
\begin{equation}
c_i = -d(\mathbf{z}_i, \boldsymbol{\mu}_{a_i}).
\label{eq:central_score}
\end{equation}
We also examine the density score to capture local compactness, defined as 
\begin{equation}
\rho_i = - \frac{1}{k} \sum_{\mathbf{z}_j \in \mathcal{N}_k(i)} d(\mathbf{z}_i, \mathbf{z}_j),
\label{eq:density_score}
\end{equation}
where $\mathcal{N}_k(i)$ represents the set of $k$ nearest neighbors within the same cluster. By ranking instances according to these metrics and selecting the top-$m$ candidates, we construct a high-confidence pool that provides stable pseudo-label supervision for subsequent reasoning and representation learning.

Across all three benchmark datasets, experimental results in Table \ref{tab:quality_score_ablation_three_datasets} demonstrate that the margin score consistently provides the most robust signal for selecting high-quality training samples. On MIntRec and MIntRec2.0, margin-based selection achieves leading average performances of 37.37 and 25.12, respectively, outperforming both density and central scores. This superiority is even more pronounced on the MELD-DA dataset, where the margin score reaches an average of 27.73, markedly surpassing the central score 24.53 and the density score 20.20. These consistent gains across diverse benchmarks suggest that prioritizing inter-cluster separability is significantly more effective for filtering noisy samples than focusing on cluster centroids or local density. The margin score's ability to identify unambiguous representatives near decision boundaries proves essential for handling complex multimodal distributions. Consequently, we adopt the margin score as our default selection criterion to ensure cleaner pseudo-labels and more stable optimization throughout the learning process.

\subsection{Effect of Confidence Ratio}
\label{app: conf}

\begin{figure}[t]
  \centering
  \includegraphics[width=\linewidth]{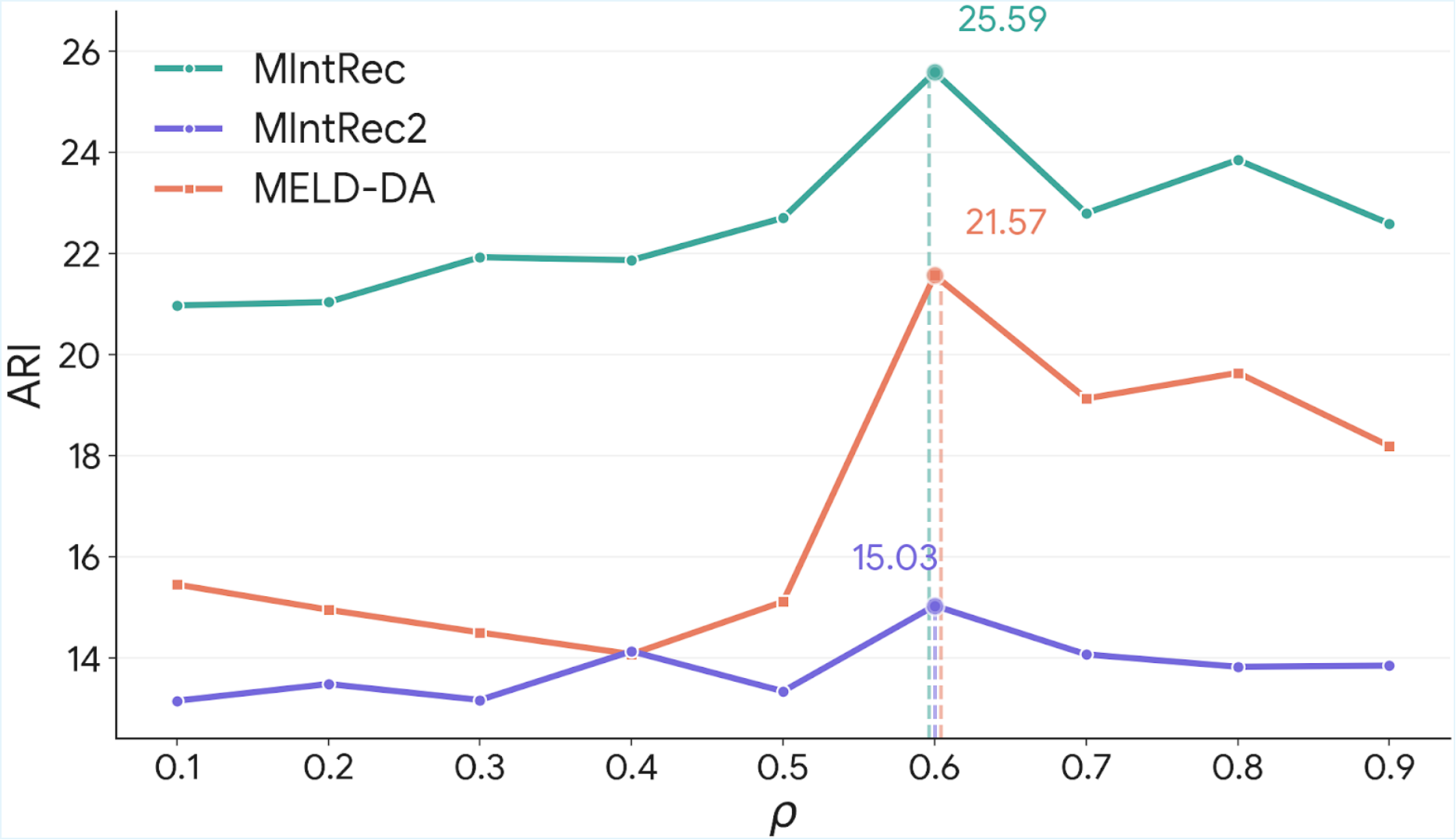}
  \caption{Effect of confidence ratio $\rho$ on the ARI metric.}
  \Description{A plot showing how the confidence ratio $\rho$ affects the ARI metric.}
  \label{fig:conf_ratio}
\end{figure}

We evaluate the impact of the high-quality confidence ratio $\rho$ by sweeping its value from 0.1 to 0.9, as illustrated in Figure \ref{fig:conf_ratio}. Across all three datasets—MIntRec, MIntRec2.0, and MELD-DA—the clustering performance consistently follows a unimodal pattern, reaching its peak at $\rho = 0.6$. Specifically, the ARI metrics achieve their maximum values of 25.59, 15.03, and 21.57, respectively. 
This consistent optimal threshold indicates that a ratio of 0.6 strikes a robust balance between pseudo-label precision and manifold coverage, regardless of the dataset's category granularity. For the fine-grained MIntRec series, this moderate ratio ensures that semantic propagation is guided by sufficiently reliable anchors, thereby preserving distinct boundaries in a dense latent space. Simultaneously, for the coarser-grained MELD-DA, this setting facilitates an extensive semantic flow that aggregates diverse samples into broader categorical structures without introducing excessive noise. 
Beyond this peak, however, the inclusion of samples with lower certainty begins to dilute the guidance provided by the semantic anchors. As $\rho$ continues to increase toward 0.9, the injection of ambiguous assignments compromises the structural integrity of the manifolds, leading to a noticeable decline in ARI across all benchmarks.

\subsection{Label-Free Temperature Selection}
\label{app:temperature_selection}

The temperatures $\tau$, $\tau_c$, and $\tau_g$ control the smoothness
of concept-supervised contrastive learning, concept-anchor assignment,
and graph propagation, respectively. To avoid tuning these parameters
with ground-truth labels, we adopt the stability-based criterion of \cite{stability}. Specifically, we evaluate all
$27$ combinations of $\tau\in\{0.3,0.5,0.7\}$,
$\tau_c\in\{0.5,1.0,1.5\}$, and $\tau_g\in\{0.5,0.7,1.0\}$ on MIntRec.
Each configuration is evaluated across five random seeds, and its
stability is computed as the average pairwise NMI between the resulting
pseudo-label assignments. This criterion requires no ground-truth labels
and favors configurations that consistently recover similar cluster
structures across runs.
Table~\ref{tab:temperature_stability} reports the top 18 configurations
among the 27 candidates. The configuration
$(\tau,\tau_c,\tau_g)=(0.3,1.5,1.0)$ achieves the highest stability of
$0.4000$, followed by $(0.3,1.0,0.5)$ and $(0.3,0.5,0.5)$. These results
demonstrate that temperature configurations can be assessed directly
from unlabeled data based on clustering consistency.

\section{Representation Visualization on MIntRec2.0 and MELD-DA
}
\label{APPD}
To qualitatively evaluate the clustering quality, this section presents a visualization analysis. We utilize t-SNE projections on the MIntRec2.0 and MELD-DA datasets to examine the resulting representation geometry and distribution patterns.

\begin{figure*}[t]
  \centering
  \includegraphics[width=\textwidth]{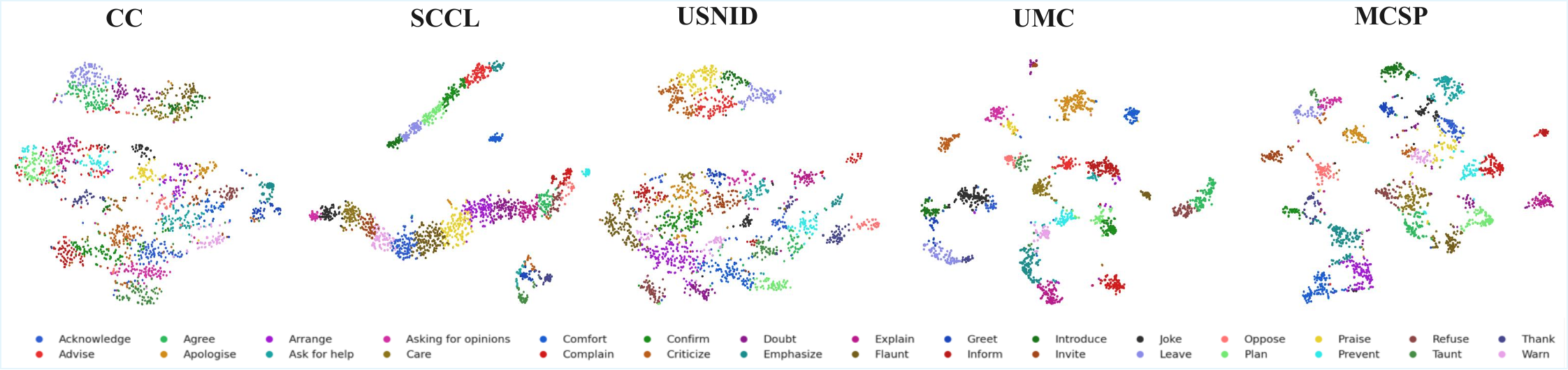}
  \caption{Visualization of representations on the MIntRec2.0 dataset.}
  \Description{A visualization of learned representations on the MIntRec2.0 dataset, showing the distribution of samples and the separability of intent clusters in the embedding space.}
  \label{fig:tsne_mintrec2}
\end{figure*}

\begin{figure*}[t]
  \centering
  \includegraphics[width=\textwidth]{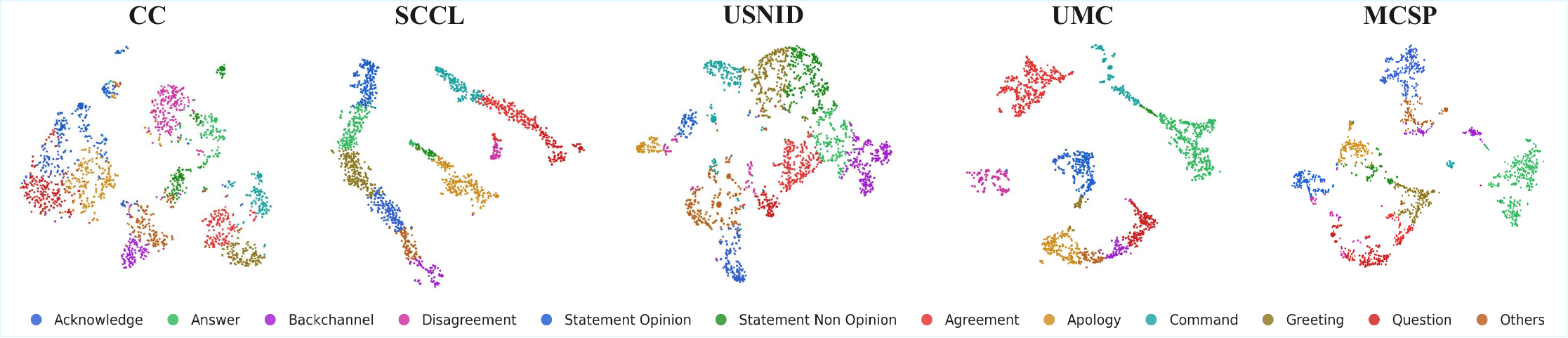}
  \caption{Visualization of representations on the MELD-DA dataset.}
  \Description{A visualization of learned representations on the MELD-DA dataset, showing the distribution of samples and the separability of intent clusters in the embedding space.}
  \label{fig:tsne_meld}
\end{figure*}



\newcommand{\mstd}[2]{#1($\pm$#2)}

Figure \ref{fig:tsne_mintrec2} and Figure \ref{fig:tsne_meld} present the t-SNE visualizations of learned representations on the MIntRec2.0 and MELD-DA datasets, respectively. On the fine-grained MIntRec2.0 dataset, all baselines exhibit significant cluster entanglement due to the increased complexity of the intent space. CC and SCCL result in large, continuous regions where multiple intents are heavily intertwined, suggesting an inability to capture subtle pragmatic distinctions. While USNID and UMC produce sharper outlines, they still suffer from partial overlaps or fragmented sub-clusters scattered across the latent space. In contrast, MCSP yields a markedly cleaner structure where most intents form single, compact, and well-separated clusters, significantly reducing ambiguity near decision boundaries.

A similar performance advantage is observed on the MELD-DA dataset. Despite having fewer categories, MELD-DA presents a challenge in maintaining intra-class cohesion across broad semantic distributions. Baselines such as CC and USNID produce diffused clusters with poorly defined boundaries, particularly for categories like Statement Opinion and Statement Non-Opinion. UMC shows improvement in cluster isolation but still exhibits scattered sample distributions. Conversely, MCSP demonstrates superior semantic aggregation, forming highly concentrated clusters that accurately reflect the underlying dialogue acts. These visualizations across both fine-grained and coarse-grained benchmarks corroborate our quantitative findings, demonstrating that MCSP consistently induces semantically coherent and geometrically well-separated intent representations regardless of the label space complexity.


\section{Analysis of Robustness}
\label{APP}
This section evaluates the robustness of MCSP from three perspectives:
the unknown category number, clustering stability across random
initializations, and the semantic consistency of generated concepts.
We first examine the practical setting where $K$ is estimated from unlabeled data. We then analyze stability of clustering performance
and generated concepts across multiple random seeds.
\subsection{Unknown Category Number}
\label{sec:unknown_k}
\begin{table}[t]
\centering
\caption{Results on MIntRec2.0 when the category number of MCSP is
estimated from unlabeled data.}
\label{tab:unknown_k}
\small
\setlength{\tabcolsep}{3.2pt}
\begin{tabular}{lccccc}
\toprule
Method & ACC & ARI & NMI & FMI & Avg. \\
\midrule
USNID
& 21.73 & 9.93 & 32.14 & 14.02 & 19.50 \\
UMC
& 27.36 & 13.50 & 36.78 & 17.34 & 23.75 \\
$\mathrm{MCSP}_{\text{Estimated }K}$
& \textbf{28.93}
& \textbf{14.19}
& \textbf{36.62}
& \textbf{18.12}
& \textbf{24.47} \\
\bottomrule
\end{tabular}
\end{table}

Following prior work, our main experiments assume that the category
number $K$ is known. This ensures consistent clustering granularity
across methods without using instance-level annotations or category
semantics. To evaluate MCSP in more practical setting, we adopt
the estimation strategy of USNID~\cite{zhang-etal-2024-clustering},
which first partitions the unlabeled pretrained representations into
an intentionally large number of clusters using K-means, discards
clusters smaller than the average cluster size, and uses the number
of retained clusters as estimated ${K}$. Across five random seeds, the estimated counts are \textbf{34, 26, 31, 28, and 31}, closely surrounding the true value of \textbf{30}. As shown
in Table~\ref{tab:unknown_k}, MCSP achieves the best average score of
$24.47\%$ with the estimated $K$, outperforming USNID and UMC by
$4.97\%$ and $0.72\%$, respectively, despite both baselines using the ground-truth
category number. These results
demonstrate that MCSP remains robust when the true category number is
unavailable.

\subsection{Stability of Intent Discovery}
\label{MCSPSTA}
\begin{table*}[t]
  \caption{Stability analysis of MCSP across five random seeds. We report mean ($\pm$ std) on each dataset.}
  \label{tab:mcsp_stability_singlecol}
  \centering
  \fontsize{9}{10.8}\selectfont
  \begin{tabularx}{\textwidth}{l@{\extracolsep{\fill}}ccccc}
    \toprule
    Dataset & ACC & ARI & NMI & FMI & Avg. \\
    \midrule
    MIntRec    & \mstd{45.21}{1.89} & \mstd{25.59}{1.62} & \mstd{48.42}{1.92} & \mstd{30.28}{1.64} & \mstd{37.37}{1.64} \\
    MIntRec2.0 & \mstd{29.19}{0.81} & \mstd{15.03}{0.55} & \mstd{37.33}{1.25} & \mstd{18.92}{0.55} & \mstd{25.12}{0.72} \\
    MELD-DA    & \mstd{34.57}{0.41} & \mstd{21.57}{1.58} & \mstd{21.61}{0.72} & \mstd{33.16}{1.55} & \mstd{27.73}{0.81} \\
    \bottomrule
  \end{tabularx}
\end{table*}

\begin{table*}[t]
  \caption{Pairwise concept stability between different random seeds on MIntRec2.0.}
  \label{tab:pairwise_stability_2x5}
  \centering
  \small
  \begin{tabular*}{\textwidth}{@{\extracolsep{\fill}}cc cc cc cc cc}
    \toprule
    \textbf{Seed Pair} & \textbf{Stability} &
    \textbf{Seed Pair} & \textbf{Stability} &
    \textbf{Seed Pair} & \textbf{Stability} &
    \textbf{Seed Pair} & \textbf{Stability} &
    \textbf{Seed Pair} & \textbf{Stability} \\
    \midrule
    (0, 1) & 0.9344 &
    (0, 2) & 0.9289 &
    (0, 3) & 0.9349 &
    (0, 4) & 0.8870 &
    (1, 2) & 0.9198 \\
    (1, 3) & 0.9291 &
    (1, 4) & 0.8913 &
    (2, 3) & \textbf{0.9399} &
    (2, 4) & 0.8927 &
    (3, 4) & 0.8910 \\
    \bottomrule
  \end{tabular*}
\end{table*}

The experimental results presented in Table \ref{tab:mcsp_stability_singlecol} demonstrate the consistent stability of the MCSP framework across five random seeds for all three benchmarks. We evaluate the performance using four standard clustering metrics alongside their average values to provide a comprehensive assessment of reliability. 

The standard deviation across most metrics remains at a notably low level, with values typically ranging from 0.41 to 1.92. On the MIntRec2.0 dataset, the fluctuations in accuracy and average performance are particularly constrained, reaching as low as 0.81 and 0.72 respectively. This high degree of consistency indicates that the proposed framework is robust and largely insensitive to the effects of random initializations. Such reliability can be attributed to the stability of our concept generation module and the structural constraints imposed by the learned manifold. Although MIntRec exhibits a relatively higher standard deviation compared to the other datasets, such as a deviation of 1.89 in accuracy, this is likely due to the inherent sensitivity of its manifold structure. In certain data distributions, the selection of representative samples can be more susceptible to minor shifts in cluster boundaries across different seeds. These subtle variations in the input context for the MLLM lead to a slightly wider range of semantic abstractions. Nevertheless, the stability across all benchmarks remains within a highly acceptable range for unsupervised discovery. These findings provide strong empirical evidence that the concepts and clusters identified by MCSP reflect intrinsic semantic structures in the multimodal data rather than artifacts of a specific random run.

\subsection{Semantic Consistency of Concepts}
\label{concept_sta}

To evaluate the robustness and reproducibility of the semantic concepts discovered by MCSP, we perform a stability analysis across five concept sets generated with random seeds 0 through 4. Stability-based evaluation is a standard criterion for assessing the reliability of unsupervised structure discovery, since it measures the consistency of outputs across repeated runs \citep{Ben-etal-2002-Stability}. Inspired by prior stability analysis of semantic components in topic models \citep{Greene-etal-2014-How}, we assess the agreement between concept sets by computing the pairwise semantic similarity of their generated concept descriptions.
Concretely, for each seed, we execute the full concept generation pipeline to obtain a set of cluster-level concepts. For each pair of seeds, we encode all generated concept phrases using a pretrained BERT model and construct a pairwise cosine similarity matrix between the two sets. Based on this similarity matrix, we align concepts via one-to-one maximum matching using the Hungarian algorithm, and define stability as the average cosine similarity of the matched pairs.


As shown in Table~\ref{tab:pairwise_stability_2x5}, MCSP achieves a high average stability of $0.9149 \pm 0.0206$, providing strong empirical evidence that the generated concepts are largely consistent across runs. This robustness primarily stems from the concept generation module, which ensures interpretability through two specific advantages. First, the extraction process is grounded on high confidence representatives. By filtering out low certainty samples, the module provides MLLMs with a pristine and noise free context, ensuring that the resulting semantic abstractions are tied to stable regions of the manifold. Second, the reasoning capability of MLLMs allows the framework to synthesize these representatives into discriminative concepts that capture the intrinsic essence of each cluster. Consequently, even when random seeds cause minor shifts in the specific samples selected as representatives, the MLLM consistently converges to the same higher order semantic interpretations. These findings confirm that MCSP concepts represent reproducible, interpretable semantic structures rather than initialization artifacts.

\section{Cost Analysis}
\label{cost}

To evaluate the efficiency of MCSP, we analyze its
end-to-end runtime and resource consumption. We first decompose the
runtime into pre-training, clustering, representative selection, MLLM
reasoning, graph construction, semantic propagation, and contrastive
learning. We then examine the resource cost of concept
generation in terms of API calls, processed video data, and token
consumption.
\subsection{Time Efficiency}
\begin{table*}[t]
  \centering
  \caption{End-to-end runtime breakdown of MCSP across three benchmarks.
  All values are reported in seconds.}
  \label{tab:time_cost_analysis}
  \scriptsize
  \setlength{\tabcolsep}{3.2pt}
  \resizebox{\textwidth}{!}{%
    \begin{tabular}{lccccccccc}
      \toprule
      Dataset
      & Total
      & Pre-training
      & Clustering
      & Rep. Selection
      & MLLM Reasoning
      & Graph Construction
      & Propagation
      & Contrastive Learning
      & Other \\
      \midrule
      MIntRec
      & 2534.16
      & 1724.96
      & 2.74
      & 0.05
      & 583.82
      & 1.54
      & 0.88
      & 81.30
      & 138.87 \\
      
      MIntRec2.0
      & 4926.59
      & 4082.00
      & 46.48
      & 1.71
      & 236.27
      & 12.21
      & 2.38
      & 296.54
      & 248.99 \\
      
      MELD-DA
      & 7779.41
      & 6709.00
      & 30.94
      & 0.44
      & 98.95
      & 13.46
      & 1.85
      & 489.97
      & 434.78 \\
      \bottomrule
    \end{tabular}%
  }
\end{table*}

We provide a detailed runtime breakdown of MCSP across the three
benchmarks, as summarized in Table~\ref{tab:time_cost_analysis}. The
reported pipeline includes pre-training, clustering, representative
selection, MLLM reasoning, graph construction, semantic propagation,
and contrastive learning. The remaining operations, including feature
extraction, anchor encoding, warm-up training, quality computation, and
bookkeeping, are grouped into the ``Other'' category.

The end-to-end runtimes are $2534.16$s, $4926.59$s, and $7779.41$s on
MIntRec, MIntRec2.0, and MELD-DA, respectively. The main computational
cost comes from standard neural training components, including
pre-training, feature extraction, and contrastive learning. For
example, pre-training takes $1724.96$s on MIntRec, while contrastive
learning takes $81.30$s, $296.54$s, and $489.97$s on the three datasets,
respectively.
By comparison, the MCSP-specific semantic propagation pipeline is
lightweight. Representative selection takes at most $1.71$s, and graph
construction together with propagation requires only $2.42$s,
$14.59$s, and $15.31$s on MIntRec, MIntRec2.0, and MELD-DA,
respectively. MLLM reasoning takes $583.82$s, $236.27$s, and $98.95$s,
but this cost is incurred only once in the offline discovery stage and
is not repeated during iterative representation refinement. These
results show that MCSP introduces limited overhead beyond conventional
multimodal training, supporting its practical use for scalable
multimodal intent discovery.

\subsection{Resource Consumption}


\begin{table}[t]
  \caption{Computational consumption of the concept generation stage across three benchmarks.}
  \label{tab:cost_analysis}
  \centering
  \resizebox{\columnwidth}{!}{%
    \begin{tabular}{lcccc}
      \toprule
      Dataset & Calls & Video Data (MB) & Input Tokens & Output Tokens \\
      \midrule
      MIntRec    & 21 & 162.68 & 16,170 & 29,150 \\
      MIntRec2.0 & 31 & 224.36 & 24,461 & 50,059 \\
      MELD-DA    & 13 & 56.49  & 9,966  & 15,896 \\
      \bottomrule
    \end{tabular}%
  }
\end{table}

We next quantify the resource consumption of the concept generation. All reasoning tasks are performed using Gemini-3.0-Pro. To limit
MLLM overhead, MCSP selects only the top three high-confidence
representatives from each cluster for semantic abstraction. It performs
one contrastive reasoning call per cluster, followed by a single global
refinement call to ensure consistency across the generated concepts.
The results in Table \ref{tab:cost_analysis} highlight the exceptional efficiency of MCSP in managing multimodal resources. Across the three benchmarks, the volume of video data actually processed remains highly condensed, ranging from 56.49 MB to 224.36 MB, with a minimal token footprint. This extreme efficiency is a direct consequence of our stratified reasoning protocol. For each dataset, MCSP requires only a single global refining call to synthesize the overall intent structure, while the remaining calls are strictly limited to contrastive reasoning for the selected representatives.

By restricting MLLM engagement to this distilled fraction of the data, the total number of API calls is kept remarkably low, such as requiring only 13 calls for the MELD-DA dataset. Detailed usage metrics further reveal that while the reasoning process consumes a certain volume of output tokens, a significant portion of the output tokens is associated with the reasoning process required for precise concept generation. The final generated concept descriptions remain remarkably concise and strictly adhere to the structural constraints defined in our prompt, as detailed in Figure \ref{fig:prompt_multimodal_contrast}.

\end{document}